\documentclass[a4paper,11pt]{article}
\pdfoutput=1 

\usepackage{jheppub} 

\usepackage[T1]{fontenc} 
\usepackage{xcolor}
\usepackage{float}
\newcommand{\ph}{\varphi}
\usepackage{mathtools,amsmath,amsthm,amssymb,physics}

\newcommand{\tiz}[1]{}
\newcommand{\gaia}[1]{}

\def\ra{\rangle}
\def\la{\langle}
\def\O{\mathcal{O}}
\def\z{\mathbf{z}}
\def\Z{\mathbb{Z}}
\def\Q{\mathbb{Q}}
\def\X{\mathcal{X}}
\def\S{\mathcal{S}}

\preprint{{\raggedleft%
ZU-TH 19/23
}}
\title{
Reduction to master integrals via intersection numbers and polynomial expansions}

\author[a]{Gaia Fontana}
\author[b]{Tiziano Peraro}

\affiliation[a]{Physik-Institut, Universit\"{a}t Z\"{u}rich, Winterthurerstrasse 190, CH-8057 Z\"{u}rich, Switzerland}
\affiliation[b]{Dipartimento di Fisica e Astronomia, Universit\'{a} di Bologna e INFN, Sezione di Bologna, via
Irnerio 46, I-40126 Bologna, Italy}

\emailAdd{gaia.fontana@physik.uzh.ch}
\emailAdd{tiziano.peraro@unibo.it}

\abstract{
Intersection numbers are rational scalar products among functions that admit suitable integral representations, such as Feynman integrals. Using these scalar products, the decomposition of Feynman integrals into a basis of linearly independent master integrals is reduced to a projection. We present a new method for computing intersection numbers that only uses rational operations and does not require any integral transformation or change of basis. We achieve this by systematically employing the polynomial series expansion, namely the expansion of functions in powers of a polynomial.  We also introduce a new prescription for choosing dual integrals, de facto removing the explicit dependence on additional analytic regulators in the computation of intersection numbers.  We describe a proof-of-concept implementation of the algorithm over  finite fields and its application to the decomposition of Feynman integrals at one and two loops.
}

\begin{document} 
\maketitle
\flushbottom

\section{Introduction}
\label{sec:intro}
Scattering amplitudes at the loop level are linear combinations of Feynman integrals.  These integrals, in dimensional regularization, obey linear relations -- most notably integration by parts identities~\cite{Tkachov:1981wb,Chetyrkin:1981qh}, Lorentz invariance identities~\cite{Gehrmann:1999as} and symmetry relations.  Using these we can reduce a large number of integrals into a much smaller set of linearly independent ones called \emph{master integrals}.  This reduction is an essential step of most modern multi-loop predictions.

The most successful and widely used integral decomposition method is the Laporta algorithm~\cite{Laporta:2000dsw}, which consists in generating and solving a large and sparse system of equations obeyed by loop integrals.  This approach produced extraordinary results and has been a key ingredient of the bulk of multi-loop predictions in the last two decades, also thanks to efficient public implementations~\cite{vonManteuffel:2012np,Smirnov:2008iw,Maierhofer:2017gsa}.  More recently, this approach has been combined with finite fields and functional reconstruction techniques~\cite{vonManteuffel:2014ixa,Peraro:2016wsq}, where complex algebraic manipulations are replaced with efficient numerical evaluations over machine-size integers and analytic results are reconstructed out of several numerical evaluations.  This method -- also implemented in both private and public~\cite{Klappert:2019emp,Smirnov:2019qkx,Peraro:2019svx,Klappert:2020nbg} programs -- significantly pushed the state of the art of multi-loop integral decomposition and, more generally, multi-loop theoretical predictions.

Despite its tremendous success, the reduction to master integrals is still a major bottleneck of higher-order theoretical predictions and it very often determines whether a multi-loop calculation is doable or not.  Moreover, this approach doesn't make manifest the vector-space structure obeyed by Feynman integrals. Indeed, since all the Feynman integrals in a family (i.e.\ having a common denominator structure) can be decomposed into a finite~\cite{Smirnov:2010hn} basis of master integrals, they can be viewed as the elements of a finite-dimensional vector space over the field of rational functions of the invariants that describe a process. In traditional reduction approaches, this property only indirectly comes out of a large number of linear identities and it is not directly exploited. The search for decomposition methods that more directly exploit this structure is definitely of interest from a purely theoretical point of view, as it is going to improve our understanding of Feynman integrals. From a more pragmatic and phenomenological viewpoint, this improved understanding may inspire the development of more efficient decomposition methods, opening up a number of important theoretical and phenomenological studies that are currently beyond our technical capability.

An approach that has been gaining a lot of interest is the study of Feynman integrals within the mathematical framework of \emph{intersection theory}~\cite{Mizera:2017rqa,Mastrolia:2018uzb,Frellesvig:2019kgj,Frellesvig:2019uqt,Mizera:2019gea,Mizera:2019vvs,Frellesvig:2020qot, Weinzierl:2020xyy,Chestnov:2022alh,Chestnov:2022xsy}.  The latter offers a method for defining rational scalar products, called \emph{intersection numbers}, in vector spaces of functions with a suitable integral representation -- among which are Feynman integrals. Using a scalar product, in turn, it is straightforward to project any element of a vector space into its components with respect to a vector basis.  Hence, the problem of decomposing any Feynman integral into a basis of master integrals is reduced to the problem of computing scalar products.

This approach, while promising, is currently not competitive with more traditional decomposition methods.  One of the main drawbacks of current approaches to the calculation of intersection numbers is the appearance of non-rational expressions in intermediate stages of the calculation.  Indeed, intersection numbers are typically computed as sums of residues of local solutions of differential equations around the poles of some rational functions.  These poles can generally appear at non-rational locations.  Because intersection numbers are rational, all non-rational contributions will eventually combine into a rational expression after summing over all residues.  The handling of non-rational expressions is however a major algebraic bottleneck.  Moreover, their appearance prevents us from using the available finite-fields-based techniques and programs in this context, since these apply to algorithms that are a sequence of rational operations.  Finally, since both the inputs and the outputs of intersection numbers are rational, it would be mathematically and theoretically more satisfactory to avoid non-rational expressions in all steps of the calculation.  A possible solution to this problem was proposed in~\cite{Weinzierl:2020xyy}, where a rational algorithm for computing intersection numbers was formulated in terms of a sequence of changes of bases, integral transformations and the application of the global residue theorem.  The required changes of bases and integral transformations are however highly non-trivial and can themselves become a bottleneck, especially for more complex integrals.

In this paper, we present a new purely rational method for computing intersection numbers which does not involve any integral transformation or change of basis.  The explicit appearance of irrational poles is sidestepped via the systematic usage of the \emph{polynomial series expansion}, or $p(z)$\emph{-adic expansion} for brevity, namely the series expansion of rational functions in powers of a prime polynomial (it is conceptually similar to $p$-adic numbers, which are instead series expansions of rational numbers in powers of a prime integer).  A $p(z)$-adic expansion is effectively equivalent to a local expansion around all the (potentially irrational) zeroes of a polynomial $p(z)$ at the same time.  However, it can simply be computed in terms of a polynomial division algorithm -- which is obviously rational -- and does not require to know the location of the roots of $p(z)$.  We thus locally solve the aforementioned differential equations in terms of $p(z)$-adic (rather than Laurent) expansions of their solution.  For the final sum over residues we then use a simple generalization of the global residue theorem that applies to functions with a known $p(z)$-adic expansion.

We implement the new method over finite fields using the \textsc{Mathematica} package \textsc{FiniteFlow} and test it on several one- and two-loop examples, finding agreement with reductions obtained with the Laporta algorithm.

Another drawback of intersection theory, when applied to loop integrals, is the need to introduce additional analytic regulators that significantly increase the complexity of the calculation and obfuscate the typical block triangular structure of integral reductions.  A mathematical formalism for dealing with this was proposed in~\cite{Caron-Huot:2021xqj,Caron-Huot:2021iev}.  In this paper we propose an alternative approach based on a suitable choice of master integrals for the dual space of Feynman integrals and on performing operations on the coefficients of the leading terms in an expansion where the additional regulators disappear.  This ``effectively'' removes any analytic dependence on the regulators in the calculation and provides a number of major simplifications, besides recovering the aforementioned block-triangular structure.

The paper is structured as follows.  In section~\ref{sec:int-th} we give a pedagogical review of intersection theory and its application to Feynman integrals, setting the notation and the main concepts used afterwards.  Section~\ref{sec:ratalg} contains the main result of this paper, namely the description of a new rational algorithm for computing intersection numbers.  In section~\ref{sec:rhotrick} we describe our new approach for dealing with analytic regulators.  An implementation of our new method over finite fields is described in section~\ref{sec:ff}.  Finally, in section~\ref{sec:ex} we validate our results with some examples and in section~\ref{sec:conclusions} we draw our conclusions and discuss possible future developments.  We give more details about the $p(z)$-adic expansion and the global residue theorem in the appendixes.

\section{Intersection numbers of Feynman integrals}
\label{sec:int-th}
In this section we set the notation and review some basic concepts of intersection theory and its application to the reduction of Feynman integrals into a basis of master integrals.  This overview is meant to be accessible to readers who are not familiar with intersection theory and we do not attempt to be mathematically rigorous or thorough. For a more comprehensive treatment of the subject and some of its applications, we refer readers to~\cite{Mizera:2017rqa,Mastrolia:2018uzb,Frellesvig:2019kgj,Frellesvig:2019uqt,Mizera:2019gea,Mizera:2019vvs,Frellesvig:2020qot,Caron-Huot:2021xqj,Caron-Huot:2021iev, Weinzierl:2020xyy,Weinzierl:2022eaz,Chestnov:2022alh, Chestnov:2022xsy, Cacciatori:2021nli,Cacciatori:2022mbi, Chen:2020uyk,Chen:2022lzr, amslaurea27132} and references therein.

We are interested in analyzing classes of integrals, such as Feynman integrals, that can be viewed as elements of a vector space.  In particular, we consider $n$-fold integrals in the variables $\z=(z_1,\ldots z_n)$ of the form\footnote{A more rigorous definition considers the elements $|\varphi_R\ra$ and $\la\varphi_L|$ as equivalence classes of $n$-forms such that forms in the same class would yield integrands that differ by total derivatives, as in eq.s~\eqref{eq:ibpclass}.  In this paper, with an abuse of notation, we identify these with the integrals themselves, since we only compute linear relations between integrals that can be written as integration by parts identities, i.e.\ in the form of eq.~\eqref{eq:ibpclass}.}
\begin{equation}
    |\varphi_R\ra = \int \dd z_1\cdots \dd z_n \, \frac{1}{u(\z)}\, \varphi_R(\z) \label{eq:phiR}
\end{equation}
and \emph{dual integrals}
\begin{equation}
    \la\varphi_L| = \int \dd z_1\cdots \dd z_n \, u(\z)\, \varphi_L(\z). \label{eq:phiL}
\end{equation}
We also refer to these as \emph{right} and \emph{left} integrals respectively. In this work, we are not concerned with studying the dependence on the integration domain, which never varies and we thus leave it implicit, with our only assumption being that integrands vanish at its boundary.  In the previous equations $\varphi_R$ and $\varphi_L$ are \emph{rational functions} in $\z$ while $u(\z)$ is a multivalued function that regulates the singularities of $\varphi_R$ and $\varphi_L$. In particular, in our applications, $u$ takes the form
\begin{equation}
    u(\z) = \prod_j B_j(\z)^{\gamma_j},
\end{equation}
where $B_j(\z)$ are polynomials and $\gamma_j$ are \emph{generic exponents}, i.e.\ exponents that are not identically equal to any integer.  For Feynman integrals, $\gamma_j$ are functions of symbolic regulators, such as the dimensional regulator. The explicit form of $u$ (or $B_j$) and the rational functions $\varphi_{R, L}$ depends on the specific problem one is interested in -- for our identifications see subsection~\ref{subsec:applic-fi}. We study classes of integrals with the same $u(\z)$ and different rational functions $\ph_{L,R}$, namely Feynman integrals within the same family.

We also assume that these (regulated) integrands vanish at the boundaries of integration.  Hence, we can write \emph{integration by parts identities (IBPs)} as
\begin{align}
    & \sum_{j=1}^n \int \dd z_1\cdots \dd z_n \, \partial_{z_j} \Big( \frac{1}{u}\, \xi^{(R)}_j \Big) ={} 0 \nonumber \\
    & \sum_{j=1}^n \int \dd z_1\cdots \dd z_n \, \partial_{z_j} \Big( u \, \xi^{(L)}_j \Big) ={} 0
\end{align}
or equivalently,
\begin{align}
   & \sum_{j=1}^n\left| \Big(\partial_{z_j} - (\partial_{z_j} u)/u \Big) \xi^{(R)}_j \right\ra = 0 \nonumber \\
    & \sum_{j=1}^n\left\la \Big(\partial_{z_j} + (\partial_{z_j} u)/u \Big) \xi^{(L)}_j \right| = 0 \label{eq:ibpclass}
\end{align}
for a list of $n$ rational functions $\xi^{(R)}_j$ and $\xi^{(L)}_j$. By generating and solving a large number of these identities, one can decompose any integral $|\varphi_R\ra$ into a \emph{basis} of linearly independent integrals $\left\{|e_j^{(R)}\ra\right\}_{j=1}^\nu$, commonly called \emph{master integrals},\footnote{We stress that, in this work, a basis of master integrals is defined to be a set of integrals that is linearly independent with respect to the set of relations in~\eqref{eq:ibpclass}. Identities which cannot be written in this form, such as some symmetry relations satisfied by Feynman integrals, are not taken into account at this stage. In practical applications, one can easily mod out results by these additional identities after the decomposition with respect to eq.s~\eqref{eq:ibpclass} has been obtained (see e.g.\ the massive two-loop sunrise example in section~\ref{sec:ex}).} where $\nu$ is the (finite) dimension of the vector space of integrals. One can similarly reduce dual integrals $\la\varphi_L|$ into a dual basis $\left\{\la e_j^{(L)}|\right\}_{j=1}^\nu$. In this work we are however interested in a more direct method for achieving such decompositions.

Our main goal is the computation of \emph{rational scalar products}
\begin{equation}
    \la \varphi_L | \varphi_R \ra
\end{equation}
between left and right integrals. These are referred to as \emph{intersection numbers}.  Intersection numbers are obviously linear in both $\varphi_L$ and $\varphi_R$ and consistent with the identities in eq.s~\eqref{eq:ibpclass}.

By calculating these scalar products and combining them, we can project left and right integrals into master integrals in the same way a vector belonging to a vector space can be decomposed into a basis via projections. Therefore we express the decomposition of any integral $|\varphi_R\ra$ and dual integral $\la\varphi_L|$ in their bases as
\begin{align}
    |\varphi_R\ra ={}&  \sum_{i=1}^{\nu} c^{(R)}_i |e_i^{(R)}\ra \nonumber \\ 
    \la \varphi_L| ={}&  \sum_{i=1}^{\nu} c^{(L)}_i \la e_i^{(L)}|
    \label{eq:generic-dec}
\end{align}
with
\begin{align}
    c^{(R)}_i ={}& \sum_{j=1}^{\nu} \qty(\mathbf{C}^{-1})_{ij} \la e_j^{(L)} | \ph_R \ra \nonumber
    \\
    c^{(L)}_i ={}& \sum_{j=1}^{\nu} \la \ph_L | e_j^{(R)} \ra \qty(\mathbf{C}^{-1})_{ji}, \label{eq:generic-dec-ccs}
\end{align}
where we introduced the \textit{metric} $\mathbf{C}_{ij}$
\begin{equation}
    \mathbf{C}_{ij} = \la e_i^{(L)} | e_j^{(R)} \ra,
\end{equation}
that contains the intersection numbers between all the master integrals.

In the following, we will thus focus on methods for computing intersection numbers and their application to the decomposition of Feynman integrals to master integrals.

\subsection{Computing intersection numbers}
\label{subsec:int-fi}
We now review a known method for computing multivariate intersection numbers~\cite{Frellesvig:2019kgj,Frellesvig:2019uqt}, which is recursive in the integration variables.  We start from the univariate case, which is the base case of the recursion, and then describe the recursive step, which computes multivariate intersection numbers of $n$-fold integrals in terms of intersection numbers of $(n-1)$-fold integrals.

While intersection numbers are rational, the method reviewed in this section generally involves non-rational intermediate expressions, namely poles of rational functions.  In section~\ref{sec:ratalg} we will propose a new method that avoids non-rational expressions in all steps of the calculation.

\subsubsection*{Univariate case}
For the univariate case, we consider intersection numbers between $1$-fold integrals of the form
\begin{align}
    |\varphi_R\ra ={}& \int \dd z\, \frac{1}{u(z)}\, \varphi_R(z) \nonumber \\
    \la\varphi_L| ={}& \int \dd z\, u(z)\, \varphi_L(z).
\end{align}
Intersection numbers are calculated as a sum of residues
\begin{equation}
    \la \ph_L | \ph_R \ra = \sum_{p\in \mathcal{P}_{\omega}} \text{Res}_{z=p}\qty(\psi\, \ph_R),\label{eq:intnum1}
\end{equation}
where the sum is extended over all $p \in \mathcal{P}_{\omega}$
\begin{equation}
    \mathcal{P}_{\omega} = \left\{ \, z \, | \, z \, \text{is a pole of} \, \omega \, \right\} \bigcup \left\{ \infty \right\}
\end{equation}
and $\psi$ is the local solution around the poles $p$ of the differential equation
\begin{equation}
    (\partial_z + \omega) \psi = \ph_L, \label{eq:diffeq1}
\end{equation}
where
\begin{equation}
    \omega \equiv \frac{\partial_z u}{u}.
\end{equation}
The local solution can be computed as a Laurent expansion around $z = p$
\begin{equation}
    \psi = \sum_{i=\text{min}}^{\text{max}} c_i \qty(z - p)^i + O((z-p)^{\text{max}+1}),
\end{equation}
which inserted in eq.~\eqref{eq:diffeq1} yields a linear system of equations for the unknown coefficients $c_i$.  The Laurent expansion can be truncated at a finite order that is sufficient for the purpose of taking the residue in eq.~\eqref{eq:intnum1}.

\subsubsection*{Multivariate case}
We now consider $n$-fold integrals of the form in eq.s~\eqref{eq:phiR} and~\eqref{eq:phiL}. We rewrite these as
\begin{align}
    |\varphi_R\ra ={}& \int \dd z_n\, |\varphi_R\ra_{n-1} \nonumber \\
    \la\varphi_L| ={}& \int \dd z_n\, \la\varphi_L|_{n-1},
\end{align}
where $|\varphi_R\ra_{n-1}$ and $\la\varphi_L|_{n-1}$ are $(n-1)$-fold integrals which have a parametric dependence on $z_n$.  Since this is a recursive method, we assume to be able to compute intersection numbers of $(n-1)$-fold integrals, which we write as
\begin{equation}
    \la \varphi_L | \varphi_R \ra_{n-1}
\end{equation}
and are \emph{rational functions} of $z_n$.
These, in turn, can be used to project any $(n-1)$-fold right and left integral into a basis of master integrals $\left\{|e_j^{(R)}\ra_{n-1}\right\}_{j=1}^{\nu_{(n-1)}}$ and $\left\{\la e_j^{(L)}|_{n-1}\right\}_{j=1}^{\nu_{(n-1)}}$ respectively,
\begin{align}
    |\varphi_R\ra_{n-1} ={}& \sum_{j=1}^{ \nu_{(n-1)}} \ph_{R,j}\, |e_j^{(R)}\ra_{n-1} \nonumber \\
    \la\varphi_L|_{n-1} ={}& \sum_{j=1}^{\nu_{(n-1)}}\, \ph_{L,j} \la e_j^{(L)}|_{n-1}, \label{eq:phinm1red}
\end{align}
where the coefficients $\varphi_{R,j}$ and $\varphi_{L,j}$ are also rational functions of $z_n$.

Intersection numbers of $n$-fold integrals can be computed as
\begin{equation}
    \la \ph_L | \ph_R \ra = \sum_{p\in \mathcal{P}_{\mathbf{\Omega}} } \text{Res}_{z_n = p}\qty( \sum_j \psi_j\, \la e_j^{(L)} | \ph_R \ra_{n-1} ), \label{eq:intnumn}
\end{equation}
where $\psi_j$ are solutions of the system of differential equations
\begin{equation}
    \partial_{z_n} \psi_j + \sum_{k=1}^{\nu_{(n-1)}}\psi_k\, \mathbf{\Omega}_{kj}= \ph_{L,j} \quad (j=1,\dots,\nu_{(n-1)})\label{eq:diffeqn},
\end{equation}
with $\ph_{L,j}$ defined as in eq.~\eqref{eq:phinm1red}, while the matrix $\mathbf{\Omega}_{jk}$ is the differential equation matrix of $(n-1)$-fold left master integrals with respect to $z_n$,
\begin{equation}
    \partial_{z_n} \la e_j^{(L)}|_{n-1} = \sum_{k=1}^{\nu_{(n-1)}}\mathbf{\Omega}_{jk}\, \la e_k^{(L)}|_{n-1}. \label{eq:Omega}
\end{equation}
In practice, one can take the derivative of the $(n-1)$-fold master integrals with respect to $z_n$ under the integral sign and decompose the resulting expression into the same master integrals to obtain the matrix $\mathbf{\Omega}_{jk}$.  The sum in eq.~\eqref{eq:intnumn} runs over the poles of the matrix elements $\mathbf{\Omega}_{jk}$
\begin{equation}
    \mathcal{P}_{\mathbf{\Omega}} = \{\, z\, | \, z \, \text{is a pole of} \, \mathbf{\Omega}
\, \} \bigcup \{ \infty \}.
\end{equation}

On a side note, we observe that eq.~\eqref{eq:intnumn} can be rewritten as
\begin{equation}
    \la \ph_L | \ph_R \ra = \sum_{p\in \mathcal{P}_{\mathbf{\Omega}} } \text{Res}_{z_n = p} \la \psi | \ph_R \ra_{n-1},
\end{equation}
where $\la \psi |_{n-1}$ solves
\begin{equation}
    \partial_{z_n} \la \psi |_{n-1} = \la \ph_{L} |_{n-1},
\end{equation}
from which the more practically useful eq.~\eqref{eq:diffeqn} is obtained by decomposing both $\la \ph_{L} |_{n-1}$ and $\la \psi |_{n-1}$ into $(n-1)$-fold left master integrals, defining $\psi_j$ as
\begin{equation}
    \la \psi |_{n-1} = \sum_{j=1}^{\nu_{(n-1)}}\, \psi_{j} \la e_j^{(L)}|_{n-1}.
\end{equation}

\subsection{Application to Feynman integrals}
\label{subsec:applic-fi}

Multiloop Feynman integrals in dimensional regularization are usually written, as a first step, in momentum representation. After a tensor decomposition, a generic $L$-loop integral of a given \emph{integral family} with $E+1$ external momenta $p_1,\ldots, p_{E+1}$ (of which only $E$ are independent because of momentum conservation) takes the form
\begin{equation}
    I[\alpha_1, \dots, \alpha_n] = \int \prod_{j=1}^L \dd^d k_j \frac{1}{z_1^{\alpha_1} \dots z_n^{\alpha_n}},\label{eq:feynint}
\end{equation}
where $z_1,\ldots,z_n$ are a complete set of denominators of loop propagators and auxiliary denominators, such that any scalar product of the form $k_i\cdot k_j$ or $k_i\cdot p_j$ is a linear combination of the $z_i$.  It follows that $n = L(L+1)/2 + LE$.  The exponents $\alpha_j$ are \emph{integers}, which can be positive, negative or  zero (negative exponents thus correspond to numerators).

Because of the vanishing of total derivatives in dimensional regularization, we can write IBP identities~\cite{Tkachov:1981wb,Chetyrkin:1981qh}
\begin{equation}
     \int \prod_{j=1}^L \dd^d k_j \pdv{}{k_i^{\mu}} \qty(\frac{v^{\mu}}{z_1^{\alpha_1} \dots z_n^{\alpha_n}})=0, \label{eq:lapibps}
\end{equation}
where $v^\mu = k_i^\mu$ or $v^\mu = p_i^\mu$.
We can exploit these identities, possibly combined with Lorentz invariance identities~\cite{Gehrmann:1999as} and symmetry relations, to perform the decomposition of a given set of Feynman integrals to master integrals. The most widely used method, i.e.\ the Laporta algorithm~\cite{Laporta:2000dsw}, consists in writing a long list of linear identities of the form of eq.~\eqref{eq:lapibps} (and similar for Lorentz invariance identities and sometimes symmetries) for different choices of $v^\mu$ and $\alpha_j$ and combining them into a large and sparse  system of linear equations satisfied by the integrals. The solution of such a system yields the reduction to master integrals (further modded out by symmetries, if included in the system).  As mentioned, solving this system is one of the main bottlenecks of many higher-order predictions.  Moreover, the vector-space structure obeyed by Feynman integrals is not manifest in this approach.

In order to work in the framework of intersection theory, we need to rewrite eq.~\eqref{eq:feynint} into a representation that mimics the one in eq.~\eqref{eq:phiR}.  Throughout this work, we use the Baikov representation~\cite{Baikov:1996iu}, which consists in changing the integration variables from loop momenta $k_j^\mu$ to propagators $z_i$. Other representations are also usable in this context, see e.g.~\cite{Lee:2013hzt,Caron-Huot:2021xqj,Caron-Huot:2021iev}. In the Baikov representation, Feynman integrals read
\begin{equation}
  I[\alpha_1, \dots, \alpha_n] = K \int \dd z_1 \dots \dd z_n \, B(z_1,\ldots,z_n)^{\gamma} \, \frac{1}{z_1^{\alpha_1} \dots z_n^{\alpha_n}} \label{eq:baikov}
\end{equation}
with $\gamma = \frac{d-L-E-1}{2}$.  The Baikov polynomial $B$ is the Gram determinant of the loop momenta and the independent external momenta, expressed as a function of the $z_i$.  $K$ is a constant that does not depend on the integration variables nor the exponents.  Feynman integrals in the Baikov representation also obey IBP identities, that are the vanishing of total derivatives with respect to the $z_j$.

Comparing eq.~\eqref{eq:baikov} to the integrals defined in section~\ref{sec:int-th}, we identify the multivalued function $u$ and the rational functions $\ph_{R,L}$ as 
\begin{equation}
    u(\z) = B^{-\gamma}\, \prod_{j=1}^n z_j^{\rho_j}\, \qquad \ph_{R,L}(\z) = \frac{1}{z_1^{\alpha_1} \dots z_n^{\alpha_n}}, \label{eq:uzfeyn}
\end{equation}
respectively.  As in~\cite{Frellesvig:2019uqt}, we inserted factors $z_j^{\rho_j}$ in $u(\z)$ because $\ph_{R,L}$ can have singularities in $z_j=0$ that are generally not regulated by the Baikov polynomial alone, while having them regulated is a requirement for applying the framework of intersection theory, as we stated in section~\ref{sec:int-th}.  Indeed, one can easily check that the differential equations we need to locally solve to compute intersection numbers have no solution around unregulated singularities.  It is understood that we are interested in the limit $\rho_j\to 0$ of a decomposition to master integrals.  We will refer to these additional variables $\rho_j$ as \emph{analytic regulators}.  Analytic regulators can generally be dropped for auxiliary denominators $z_i$ which only appear with a negative exponent in $\ph_{R,L}$, if the corresponding poles at infinity are regulated by the Baikov polynomial (which is usually the case).

The need of analytic regulators is a further complication of the application of intersection theory to Feynman integrals, since it adds additional parameters into the problem and obfuscates the typical block triangular structure of integral reduction identities.  We will show in section~\ref{sec:rhotrick}, however, that if we are interested in the decomposition of Feynman integrals, we can effectively remove the explicit dependence on $\rho_j$ from the calculation of intersection numbers, by making a suitable choice of basis for dual integrals and systematically work in the $\rho_j\to 0$ limit. An alternative mathematical framework for dual Feynman integrals which also removes the need of regulators was proposed in~\cite{Caron-Huot:2021xqj,Caron-Huot:2021iev}.

In this paper we identify Feynman integrals in the Baikov representation as right integrals $|\ph_R\ra$ and their duals as left integrals $\la \ph_L|$.  We therefore apply the framework of intersection theory for computing intersection numbers between Feynman integrals and their duals. This approach allows to directly decompose an integral into the basis of chosen master integrals via the projections in~\eqref{eq:generic-dec}, without resorting to the construction of an IBP system. For a given integral, this approach implies the necessity to calculate intersection numbers of $n$-fold integrals, with $n$ being the number of independent denominators.

\section{Intersection numbers via polynomial series expansions}
\label{sec:ratalg}

In this section, we present the main result of this paper, namely a new rational method for computing intersection numbers via the systematic use of \emph{polynomial series expansions}, or $p(z)$-adic expansions for brevity.

As mentioned, one of the main drawbacks of the method described in section~\ref{sec:int-th} is the appearance of non-rational poles in the calculation of intersection numbers.  While intersection numbers are rational, irrational intermediate expressions represent a bottleneck in complicated algebraic manipulations and make the usage of efficient techniques based on finite fields impractical.

Because we always deal with rational integrands, all their (rational and non-rational) poles are roots of rational polynomial factors of their denominators. In the following, we will thus consider each of these polynomial factors, which are \emph{prime}\footnote{We can, in principle, relax the requirement of having prime polynomial factors to the one of having a list of pairwise co-prime polynomial factors.} over $\Q$, rather than individual poles, as the building blocks of our calculation.

\subsection{$p(z)$-adic series expansions}
\label{sec:pzadic}
We elaborate on the expansion of any rational function in terms of a polynomial $p(z)$, a necessary step to calculate intersection numbers with our newly-proposed rational algorithm.

Consider a univariate polynomial $p(z)$ of degree $\deg p$ in the variable $z$ that is prime over the field of rational numbers. We define the $p(z)$-adic series expansion of a rational function $f(z)$ as
\begin{equation}
    f(z) = \sum_{i=\textrm{min}}^{\textrm{max}} c_i(z)\, p^i(z) +\O\left(p(z)^{\textrm{max}+1}\right),\label{eq:pzadic}
\end{equation}
where the coefficients $c_i(z)$ are polynomials in $z$ over the rational field, whose degree is lower than the one of $p(z)$
\begin{equation}
    c_i(z) = \sum_{j=0}^{\deg p-1} c_{ij}z^j, \quad c_{ij} \in \Q.\label{eq:pzadiccz}
\end{equation}
The notation $\O((p(z)^k)$ stands for terms proportional to the $k$-th power of the polynomial $p(z)$.  If $z$ is such that $p(z)=\delta$, where $\delta$ is some small quantity, then we obviously have $\O(p(z)^k)=\O(\delta^k)$.  Hence, the $p(z)$-adic expansion allows us to consider the limit $p(z)\to 0$, without resorting to non-rational operations nor knowing the explicit location of the (irrational) roots of $p(z)$.  In particular we have
\begin{equation}
    f(z)|_{p(z)=\delta} =  \mathcal{O}(\delta^\textrm{min}),
\end{equation}
where $i=\textrm{min}$ is the starting order of the expansion.  Similarly to Laurent expansions, the latter can be deduced from polynomial factors of the form $p(z)^k$ in $f(z)$, but unlike Laurent expansions around irrational points, here they can be detected using only operations over the rational field.

When calculating intersection numbers, we are often required to expand around the irrational poles of $\omega$ or $\mathbf{\Omega}$. These irrational poles are the roots of some polynomials $p(z)$ irreducible over $\Q$. Therefore a $p(z)$-adic expansion is effectively equivalent to considering all expansions of a function around all the roots of the polynomial at once, ultimately avoiding their explicit appearance.

The expansion in~\eqref{eq:pzadic} can be obtained by repeated polynomial divisions modulo $p(z)$ with remainder, as illustrated in appendix~\ref{sec:polydiv}.
As a simpler alternative, we observe that we can implement the same expansion via a shortcut that avoids repeated polynomial divisions.  We introduce an auxiliary parameter $\delta$ and define, for a generic function $f(z)$, its remainder\footnote{As explained in more detail in appendix~\ref{sec:polydiv}, this also involves computing the multiplicative inverse of the denominator of $f(z)$ modulo $p(z)-\delta$. Some computer algebra systems have builtin functions for polynomial reminders that are also applicable to rational functions, such as the \texttt{PolynomialRemainder} function in \textsc{Mathematica}.} modulo $p(z)-\delta$
\begin{equation}
    \lfloor f(z) \rfloor_{p(z)-\delta} \equiv f(z) \, \text{mod}\, \big( p(z)-\delta \big),
\end{equation}
that is a polynomial of degree $\deg p - 1$ in $z$ and has a rational dependence on the parameter $\delta$.
We highlight that this step amounts to the substitution $p(z)=\delta$ and is exact, i.e.\ we obtain back the original function by substituting $\delta\to p(z)$ in the right-hand-side.  After doing this, we series expand for small $\delta$, obtaining
\begin{equation}
    \lfloor f(z) \rfloor_{p(z)-\delta}\eval_{\delta\rightarrow 0} = \sum_{i=\textrm{min}}^{\text{max}} \sum_{j=0}^{\deg p-1} c_{ij} z^j \delta^i + O(\delta^{\text{max}+1}),
\end{equation}
where the coefficients $c_{ij}$ are identified with those appearing in the $p(z)$-adic expansion as in eq.s~\eqref{eq:pzadic} and~\eqref{eq:pzadiccz}.
In this way we obtain the $p(z)$-adic expansion of a rational function up to the order $O(p(z)^{\text{max}+1})$.

Polynomial series expansions can be combined with the \emph{univariate global residue theorem} (see appendix~\ref{sec:globres}) to compute the sum of the residues of a rational function around the poles of a rational polynomial, with purely rational operations.  A polynomial $p(z)$ can always be factorized over the complex field as
\begin{equation}
    p(z) = l_c\, \prod_{k=1}^{\textrm{deg } p}(z-y_k),
\end{equation}
where $l_c$ is the \emph{leading coefficient}, i.e.\ the coefficient of $z^{\textrm{deg } p}$ in $p(z)$, and $y_k$ are the roots of $p(z)$.
For a generic function $f(z)$ we define
\begin{equation}
\text{Res}_{p(z)} \qty( f(z) ) \equiv 
    \sum_{k=1}^{\textrm{deg }p} \text{Res}_{z=y_k} \qty( f(z) ).\label{eq:respzdef}
\end{equation}
i.e.\ the sum of the residues of $f(z)$ at the roots of $p(z)$.
If $f(z)$ admits the $p(z)$-adic expansion in eq.~\eqref{eq:pzadic}, then the sum of residues in the previous equation can be computed without resorting to irrational operations and without explicitly knowing the location of the roots $y_k$, using the following generalization of the global residue theorem
\begin{equation}
    \text{Res}_{p(z)} \qty( f(z) ) = \frac{c_{-1, \textrm{deg } p - 1}}{l_c},
    \label{eq:respz}
\end{equation}
where the coefficients $c_{ij}$ are defined as in eq.~\eqref{eq:pzadiccz}.

We now apply $p(z)$-adic expansions to the computation of univariate and multivariate intersection numbers.

\subsection{Univariate intersection numbers}
For the computation of intersection numbers of univariate integrals in the framework of $p(z)$-adic expansion, we rewrite eq.~\eqref{eq:intnum1} as
\begin{equation}
    \la \ph_L | \ph_R \ra = \sum_{p(z)\in \mathcal{P}_{\omega}[z]} \la \ph_L | \ph_R \ra_{p(z)},\label{eq:intnum1pz}
\end{equation}
where the sum runs over all factors $p(z)$ of the denominators of $\omega$
\begin{equation}
    \mathcal{P}_{\omega}[z] = \left\{\textrm{factors of the denominator of $\omega$} \right\} \bigcup \left\{ \infty \right\}. \label{eq:Pz}
\end{equation}
The contribution $\la \ph_L | \ph_R \ra_{\infty}$ is computed exactly as the contribution at $p=\infty$ to eq.~\eqref{eq:intnum1}.  Each other addend of the form \textbf{$\la \ph_L | \ph_R \ra_{p(z)}$} is instead the sum of all contributions to~\eqref{eq:intnum1} coming from the roots of~$p(z)$.

For each factor $p(z)$, we make an ansatz for the local solution $\psi$ of the differential equation~\eqref{eq:diffeq1} close to all the roots of $p(z)$.  This takes the form of a $p(z)$-adic expansion  
\begin{equation}
    \psi  = \sum_{i=\text{min}}^{\text{max}} \sum_{j=0}^{\deg p-1} c_{ij}z^j p(z)^i + O(p(z)^{\text{max}+1}).\label{eq:psipz}
\end{equation}
We stress that, even when $p(z)$ has irrational roots, these do not explicitly appear in eq.~\eqref{eq:psipz} which, once plugged into eq.~\eqref{eq:diffeq1} and $p(z)$-adic expanded again, yields a linear system of equations with rational coefficients that is then solved for the unknowns $c_{ij}$.

The solution for $\psi$ is then multiplied by $\ph_R$ and the product is again $p(z)$-adic expanded with respect to $p(z)$. We obtain
\begin{equation}
    \psi\, \ph_R  = \sum_{i}\sum_{j=0}^{\deg p-1} \tilde{c}_{ij} z^j p(z)^i + O(p(z)^{0})
\end{equation}
and apply the univariate global residue theorem as in eq.~\eqref{eq:respz}
\begin{equation}
    \la \ph_L | \ph_R \ra_{p(z)} = \frac{\tilde{c}_{-1,\deg p-1}}{l_c},
\end{equation}
where $l_c$ is the leading coefficient of $p(z)$.
As we stated, this is the sum of the contributions to the intersection number between $\la \ph_L | \ph_R \ra$ given by all the roots of $p(z)$.

\subsection{Multivariate intersection numbers}
\label{sec:multivpz}
The generalization to the multivariate case is straightforward.  In this subsection we let $z\equiv z_n$.  We rewrite eq.~\eqref{eq:intnumn} as
\begin{equation}
    \la \ph_L | \ph_R \ra = \sum_{p(z)\in \mathcal{P}_{\mathbf{\Omega}}[z]} \la \ph_L | \ph_R \ra_{p(z)},\label{eq:intnumnpz}
\end{equation}
where the sum is over the factors $p(z)$ of the denominators of $\mathbf{\Omega}$.
\begin{equation}
    \mathcal{P}_{\mathbf{\Omega}}[z] = \left\{\textrm{factors of the denominator of $\mathbf{\Omega}_{ij}$} \right\} \bigcup \left\{ \infty \right\}. \label{eq:Pzn}
\end{equation}
As before, the contribution $\la \ph_L | \ph_R \ra_{\infty}$ is computed exactly as the contribution at $p=\infty$ to eq.~\eqref{eq:intnumn}.  Each other addend of the form \textbf{$\la \ph_L | \ph_R \ra_{p(z)}$} is instead the sum of all contributions to~\eqref{eq:intnumn} coming from the roots of~$p(z)$.

We make an ansatz for the local solutions $\psi_i$ to the differential equation~\eqref{eq:diffeqn} around all the roots of $p(z)$ that, as in the univariate case, take the form of $p(z)$-adic expansions
\begin{equation}
    \psi_i  = \sum_{j=\text{min}}^{\text{max}} \sum_{k=0}^{\deg p-1} c_{ijk}z^k p(z)^j + O(p(z)^{\text{max}+1}) \quad (i=1,\dots,\nu_{n-1}).
\end{equation}Eq.~\eqref{eq:diffeqn} thus yields a linear system that we solve for the unknowns $c_{ijk}$.

We then take the scalar product of $\psi_j$ and $\la e_j^{(L)} | \ph_R \ra_{n-1}$ and $p(z)$-adic expand with respect to $p(z)$. We obtain
\begin{equation}
    \sum_{i=1}^{\nu_{(n-1)}} \psi_i\, \la e_i^{(L)} | \ph_R \ra_{n-1}  = \sum_{j}\sum_{k=0}^{\deg p-1} \tilde{c}_{jk} z^k p(z)^j + O(p(z)^{0}).
\end{equation}
The contribution of the roots of $p(z)$ to the intersection number $\la \ph_L | \ph_R \ra$ is again given by
\begin{equation}
    \la \ph_L | \ph_R \ra_{p(z)} = \frac{\tilde{c}_{-1,\deg p-1}}{l_c}.
\end{equation}

\section{Dual integrals and analytic regulators}
\label{sec:rhotrick}
As explained in subsection~\ref{subsec:applic-fi}, one of the drawbacks of the application of intersection theory to Feynman integrals is the need of introducing \emph{analytic regulators} $\rho_j$ for each integration variable $z_i$ that may appear as a denominator for $\ph_R$ or $\ph_L$.  However, since we are interested in computing intersection numbers for the purpose of reducing Feynman integrals~$\ph_{R}$ to master integrals, we can exploit the freedom of choice for the dual basis to simplify our problem.  Indeed, while intersection numbers obviously depend on the dual integrals, the coefficients of the decomposition of a Feynman integral~$|\varphi_R\ra$ are independent of the choice of dual bases.  In this section we discuss a simple strategy that ``effectively'' removes the explicit dependence on analytic regulators from the calculation, without substantially changing the algorithm for computing intersection numbers.  An alternative mathematical formalism for choosing dual integrals and dealing with this issue was presented in~\cite{Caron-Huot:2021xqj,Caron-Huot:2021iev}.  A comparison between the two approaches, as well as an in-depth study of all implications and possible limitations of our new approach, is outside the scope of this work and is left to future investigations.

\subsection{Choice of dual integrals}

Our approach simply consists in choosing dual integrals of the form
\begin{equation}
    \ph_{L}(\z) = \rho_1^{\Theta(\alpha_1-\frac{1}{2})}\cdots \rho_n^{\Theta(\alpha_n-\frac{1}{2})}\frac{1}{z_1^{\alpha_1} \dots z_n^{\alpha_n}} \label{eq:rhophil}
\end{equation}
and \emph{systematically} work in the $\rho_j\to 0$ limit.  In other words, we multiply each factor $z_j^{-\alpha_j}$ by the analytic regulator $\rho_j$ if (and only if) $\alpha_j>0$, i.e.\ when $z_j$ appears in the denominator of $\ph_L$.  Then, each step in the algorithm is worked out by keeping only the leading contributions in the $\rho_j\to 0$ limit.

One can easily see that the dependence on $\rho_j$ of intersection numbers only appears at the $j$-th level of the recursion.  On the other hand, intersection numbers of $j$-fold integrals are systematically computed in the $\rho_j\to 0$ limit, which is \emph{finite} when we use left integrands of the form~\eqref{eq:rhophil}.\footnote{This is true if at most single poles in each regulator $\rho_j$ appear when solving the differential equations required for computing intersection numbers.  This can be easily proved to be true in the univariate case. In the multivariate case, we empirically find it to be true in all the examples we computed, but exceptions may exists and may be investigated in future works.}  Hence, we only have to consider one regulator at a time, namely $\rho_j$ (and only $\rho_j$) at the $j$-th step of the recursion.

Consider a generic integration variable $z=z_j$ and the corresponding regulator $\rho=\rho_j$.  When solving eq.~\eqref{eq:diffeq1} for $\psi$, for each order of the Laurent or $p(z)$-adic series expansion in $z$ of the equation, we only consider the leading term in a $\rho\to 0$ expansion (similar statements hold for $\psi_j$ solving eq.~\eqref{eq:diffeqn}).  Since we systematically work on the leading coefficients of a $\rho\to 0$ expansion and obtain \emph{finite}  intersection numbers after each step of the recursion, the calculation is effectively as simple as one without any dependence on $\rho$ -- or even simpler in most cases.

For a more detailed discussion, we distinguish the two cases of exponents $\alpha\leq 0$ and exponents $\alpha > 0$ inside $\ph_L \sim z^{-\alpha}$.  If $\alpha\leq 0$, then $\ph_L$ has no singularity in $z=0$ hence we do not expect to need an analytic regulator at all and one may set $\rho=0$ before solving for $\psi$.  If instead $\alpha > 0$, since poles in $z=0$ are generally \emph{not} regulated by $\omega$, eq.~\eqref{eq:diffeq1} has no solution unless we add an analytic regulator.  The solution for $\psi$, in turn, develops a pole in $\rho=0$, which cancels against the prefactor $\rho$ we add according to the prescription in~\eqref{eq:rhophil}, yielding a finite solution for $\psi$ in the $\rho\to 0$ limit.  We observe that, therefore, for $\alpha > 0$ the solution for $\psi$ is only non-zero (in the $\rho\to 0$ limit) around poles at $z=0$ and possibly at $z=\infty$, hence we can skip the calculation at all other poles or denominator factors.

This strategy has a number of advantages.
\begin{itemize}
    \item The calculation is \emph{effectively} independent of $\rho_j$ since we are directly working on the leading coefficients of expansion around $\rho_j\to 0$.  The full dependence on $\rho_j$ is never considered and, when we implement the algorithm over finite fields, we do not sample over or reconstruct the $\rho_j$ dependence.
    \item Intermediate analytic expressions (if reconstructed) are dramatically simpler.
    \item The matrix of intersection numbers $\la \ph_L | \ph_R \ra$ for a list of $\ph_L$ and $\ph_R$, as well as the matrices $\mathbf{C}_{ij}$ and $\mathbf{\Omega}_{ij}$ have a block triangular structure, similar to what we expect from integral decompositions in traditional approaches, where blocks correspond to so-called \emph{sectors}.  In particular, $\la \ph_L | \ph_R \ra \neq 0$ only if the set of indexes $k$ such that $\alpha_k>0$ for the integrand $\ph_L$ is either the same set or a subset of the one for $\ph_R$ -- in the language of Feynman integrals, we say that $\ph_L$ belongs to a sector that is a subsector of $\ph_R$.  The same structure appears in the matrices $\mathbf{C}_{ij}$ and $\mathbf{\Omega}_{ij}$ (examples are given in section~\ref{sect:examples}).  We note that this structure is transposed to the one that is observed in the reduction or differential equations for (right) Feynman integrals $|\ph_R\ra$.
    \item A large set of intersection numbers and contributions of poles to intersection numbers vanishes, for the reasons illustrated above.
    \item At intermediate steps of the recursive algorithm, sometimes there are \emph{fewer master integrals} with respect to the case where the full $\rho$ dependence is considered.
\end{itemize}

\subsection{Subtleties of analytic regulators}
\label{sec:subtlerho}

A few subtleties need to be discussed.  The system of equations for the coefficients of the expansion of $\psi$ in $z$ can sometimes be underdetermined, regardless of the sign of the exponent $\alpha$ of $z$.  In this case, we add more constraints coming from higher-order terms in the $\rho\to 0$ expansion of the equations.  This can also happen when $\alpha\leq 0$ -- in this case a regulator is temporarily added -- as well as for contributions to intersection numbers from denominator factors $p(z)\neq z,\infty$ (see eq.\eqref{eq:intnumnpz} and~\eqref{eq:Pzn}) where we can regulate the system with the prescription
\begin{equation}
    \omega \to \omega + \rho_p\frac{\partial_z\, p(z)}{p(z)} \qquad \textrm{or} \qquad \mathbf{\Omega}_{ij} \to \mathbf{\Omega}_{ij} + \rho_p\frac{\partial_z\, p(z)}{p(z)}\, \delta_{ij} .\label{re:extrarho}
\end{equation}
Even in such cases, we still solve the equation for the coefficients of $\psi$ in the $\rho_p\to 0$ limit, keeping higher-order terms in such limit only as necessary.  In most cases the additional constraints simply amount to setting to zero the previously undetermined coefficients, hence the next-to-leading equations in the $\rho\to 0$ (or $\rho_p\to 0$) expansion can almost always be dropped after a numerical check.

Another subtlety is that, even if a variable $z$ is not a factor of the Baikov polynomial, it is still possible that $\mathbf{\Omega}$ may have $z$ as a denominator factor at the corresponding step of the recursion.  This, in turn, might regulate the singularity at $z \to 0$ for at least some of the integrals $\ph_L$, without the need of the corresponding analytic regulator $\rho$.  In such cases $\psi$ does not develop the singularity $1/\rho$ that cancels the $\rho$ prefactor in eq.~\eqref{eq:rhophil} and therefore we obtain $\psi=0$ for all poles -- which for the purpose of computing intersection numbers is equivalent to setting $\la \ph_L|=0$.  While in principle we may remove the $\rho$ prefactor from eq.~\eqref{eq:rhophil} when this happens, in all our tests this was not needed.  Indeed, even in such cases we have always found enough integrals of the form in~\eqref{eq:rhophil} to obtain a complete basis of dual integrals in the $\rho\to 0$ limit.

As a final remark, our approach is only meant to deal with singularities at $z=0$ or underdetermined systems for expansions around arbitrary factors $p(z)$.  In cases where a system with no solution appears for a $p(z)\neq z,\infty$, both this approach and the traditional one that considers the full dependence of intersection numbers on the analytic regulators in~\eqref{eq:uzfeyn} are problematic and deserve further investigation.  In principle the prescription in~\eqref{re:extrarho} may fix the singular point, but the solution for $\psi$ generally develops a singularity in $\rho_p=0$ which only cancels after the intersection numbers are combined into a decomposition as in eq.~\eqref{eq:generic-dec-ccs}.  It is possible that a generalization of the strategy proposed here -- namely multiplying the offending integrands by $\rho_p$ and systematically work in the $\rho_p\to 0$ limit after applying eq.~\eqref{re:extrarho} -- may also simplify the calculation in these cases.

\section{Implementation over finite-fields}
\label{sec:ff}
One of the main motivations for developing a rational algorithm for intersection numbers, besides avoiding algebraic difficulties in handling non-rational roots, is the possibility of implementing it over finite fields. Finite fields are numerical fields with a finite number of elements. In the context of scattering amplitudes, the most common choice is the field of integers modulo a prime number, namely $\Z_p=\left\{0,\ldots,p-1\right\}$ with $p$ a prime.  In $\Z_p$ we can perform efficient numerical evaluations using machine-size integers (for primes $p<2^{64}$) which are also exact, i.e.\ not affected by numerical inaccuracies.  We can implement over $\Z_p$ any algorithm that is rational, i.e.\ a sequence of rational arithmetic operations.  Out of these numerical evaluations, we can thus use rational reconstruction techniques to infer full analytic results as rational functions of the free parameters of the problem (see e.g. \ ~\cite{vonManteuffel:2014ixa,Peraro:2016wsq,Klappert:2019emp,Peraro:2019svx} for more details).

In this section we describe a proof-of-concept implementation of our new algorithm over finite fields that uses the \textsc{Mathematica} package \textsc{FiniteFlow}~\cite{Peraro:2019svx}.  This package offers the possibility of building new algorithms by combining core algorithms in computational graphs (dataflow graphs) from a high-level interface and reconstruct analytic results out of numerical evaluations.

We give an overview of how our implementation of multivariate intersection numbers works.  At first, for simplicity, we will assume to know in advance a basis of master integrals and dual master integrals for all levels of the recursive algorithm.  We will then address this point more thoroughly in subsection~\ref{sec:mis}.

\subsection{Setting up the recursive calculation}
Before doing any explicit calculation, we make a list of intersection numbers we need to compute at each step of the recursive algorithm.  Our starting point is a list of $n$-variate intersection numbers we wish to compute.  This will typically include all the intersection numbers needed for the decomposition of a selection of integrals $|\ph_R\ra$ via eq.s~\eqref{eq:generic-dec} and \eqref{eq:generic-dec-ccs}, namely $\la e^{(L)}_j|\ph_R\ra$ and the ones appearing in the metric $\mathbf{C}_{ij}=\la e^{(L)}_i| e^{(R)}_j\ra$, for a basis $\left\{|e_j^{(R)}\ra\right\}_{j=1}^\nu$ and a dual basis $\left\{\la e_j^{(L)}|\right\}_{j=1}^\nu$ -- but we consider a generic list of $n$-variate intersection numbers $\la \ph_L|\ph_R\ra$ for the sake of generality.  From these, by reviewing the algorithm in subsection~\ref{subsec:int-fi}, we can easily make a list of $(n-1)$-variate intersection numbers which are required as input for the recursive algorithm.  More precisely, if $\left\{|e_j^{(R)}\ra_{n-1}\right\}_{j=1}^{\nu_{(n-1)}}$ and $\left\{\la e_j^{(L)}|_{n-1}\right\}_{j=1}^{\nu_{(n-1)}}$ are the chosen bases for right and left $(n-1)$-fold integrals respectively, the calculation of the $n$-variate intersection numbers in the list requires the following $(n-1)$-variate intersection numbers
\begin{itemize}
    \item the matrix elements of the $(n-1)$-variate metric $\la e^{(L)}_i| e^{(R)}_j\ra_{n-1}$
    \item $\la \ph_L | e^{(R)}_j \ra_{n-1}$ for the decomposition of $\la \ph_L|$ needed in  the differential equation~\eqref{eq:diffeqn}
    \item $\la d^{(L)}_i| e^{(R)}_j\ra_{n-1}$ with $\la d^{(L)}_i|_{n-1}\equiv \partial_{z_n}\la e^{(L)}_i|_{n-1}$ needed for computing the matrix elements $\mathbf{\Omega}_{ij}$ defined in eq.~\eqref{eq:Omega}
    \item $\la e^{(L)}_j | \ph_R \ra_{n-1}$ appearing in eq.~\eqref{eq:intnumn}.
\end{itemize}
This makes up a new list of required $(n-1)$-variate intersection numbers.  We proceed recursively, by applying the same analysis to this new list, obtaining all the required $(n-2)$-variate intersection numbers, until we reach the univariate case.

Using this method, we easily generate a list of required intersection numbers for each step of the recursive algorithm, before any explicit calculation is done.  We can thus proceed with the implementation of the algorithm described in section~\ref{sec:ratalg}, starting from univariate intersection numbers and moving up until the $n$-variate case.

\subsection{Laurent and $p(z)$-adic expansions}
In this subsection, we describe how we implement Laurent and $p(z)$-adic expansions over finite fields, which are extensively used in all steps of the recursion.  While \textsc{FiniteFlow} already has algorithms for reconstructing Laurent expansions, we find it is more convenient to adopt a different method that uses the sparse linear solvers implemented in the package.

The starting point is a rational function $f(z)$ for which we want to compute the series expansion, which we write as
\begin{equation}
    f(z) = \frac{n(z)}{d(z)} = \frac{\sum_j n_j\, z^j}{\sum_{j} d_j z^j}
\end{equation}
and for which we have reconstructed the full dependence on $z$, while the coefficients $n_j$ and $d_j$ are generally just known numerically, i.e.\ they can be evaluated via a rational algorithm.

A Laurent expansion for $f(z)$ around $z=0$ can be easily obtained by making an ansatz for it
\begin{equation}
    f(z) = \sum_{j=\min}^{\max} c_j \, z^j + \O(z^{\max +1}), \label{eq:lauransatz}
\end{equation}
where $\min$ is easily determined from the lowest-degree non-vanishing terms in the denominator and numerator of $f$, while $\max$ is the order at which the expansion is needed.  We thus plug eq.~\ref{eq:lauransatz} into the relation
\begin{equation}
    f(z)\, d(z) - n(z) = 0, \label{eq:fdmn0}
\end{equation}
truncated at the suitable order determined by $\max$, and impose that the coefficient of all terms proportional to $z^k$ in the previous equation must vanish.  This gives a linear system of equations for the coefficients $c_i$ in~eq.~\ref{eq:lauransatz}.  It is easy to see that this system is triangular, hence equivalent to a list of substitutions.  Expansions around $z=\infty$ can be obtained similarly, by considering instead the function
\begin{equation}
    g(\delta) = -\frac{1}{\delta^2}\, f\left(\frac{1}{\delta}\right)
\end{equation}
where the prefactor $-1/\delta^2$ is the Jacobian of the transformation.  Hence we expand $g(\delta)$ for $\delta \to 0$, using the same method illustrated above.  Note that the coefficients in the numerator and the denominator of $g$ coincide with the ones for $f$, except that they multiply different monomials, hence there's no need of shifting variables in this case.

The calculation of $p(z)$-adic expansions is instead done in two steps.  Let $p(z)$ be the polynomial with respect to which we want to expand our functions
\begin{equation}
    p(z) = \sum_{k=0}^{\deg p} a_k\, z^k.
\end{equation}
In the first step we compute
\begin{equation}
    z^k \mod \big( p(z)-\delta\big)
\end{equation}
for all needed values of the exponent $k$.
This is done by writing a linear system of equations of the form
\begin{equation}
    \sum_{k=0}^{\deg p} a_k\, z^{k+\alpha}\, \delta^{\beta} - \delta^{\beta+1} = 0,
\end{equation}
for integers $\alpha$ and $\beta$, where the unknowns are monomials $z^k \delta^j$. We solve these linear relations for the monomials, expressing monomials with higher powers in $z$ as linear combinations of monomials with lower powers in $z$ (and higher powers in $\delta$).  This yields the $p(z)$-adic expansion of every monomial we need, that takes the form
\begin{equation}
    z^k = \sum_{j} \sum_{l=0}^{\deg p -1} b_{kjl} z^l\, \delta^j . \label{eq:zkred}
\end{equation}
We stress that this expansion is exact, i.e.\ it will produce an exact identity after substituting $\delta\to p(z)$, although in practice it can also be truncated to the needed order in $\delta$.  We also notice that the system is triangular, hence equivalent to a sequence of linear substitutions.  In the second step, we proceed similarly to the expansion around $z=0$, by making an ansatz for the $p(z)$-adic expansion of $f$
\begin{equation}
    f(z) = \sum_{j=\min}^{\max} \sum_{k=0}^{\deg p - 1} c_{jk} z^k\, \delta^j + \O(\delta^{\max +1}),
\end{equation}
where we have set $p(z)=\delta$ for convenience.  The minimum order can be inferred by the order of the expansion of the denominator $d(z)$, which in turn can be easily derived by substituting eq.~\eqref{eq:zkred} into its expression. We thus substitute this expansion into eq.~\eqref{eq:fdmn0}, perform the substitutions in~\eqref{eq:zkred} and impose that all the coefficients of terms $z^k\, \delta^j$ of the (appropriately truncated) expansion of~\eqref{eq:zkred} vanish.  This yields a linear system of equations for the coefficients $c_{jk}$. Notice that, in the special case where $p(z)$ is linear, eq.~\eqref{eq:zkred} is equivalent to a shift of variables and the second step becomes identical to the expansion around zero discussed above.

We observe that we typically need to expand several rational functions having the same denominator.  As a shortcut, we expand a generic numerator where the coefficients $n_j$ are symbolic.  In practice, we treat $n_j$ as additional unknowns in our linear system, so that the solution for the coefficients of the expansion will be a linear combination of the coefficients $n_j$.  Moreover, for expansions around $z=0$ and $z=\infty$, we remove factors of $z$ in the denominator by allowing the numerator $n(z)$ to contain terms with negative powers of $z$.  This allows us to combine the expansion of a larger set of functions together.

\subsection{Polynomial factors and multivariate intersection numbers}
We now describe the most important step of the implementation, namely the recursive step in the calculation of multivariate intersection numbers.  Both the univariate and the multivariate algorithm illustrated in section~\ref{sec:ratalg} consist in the computation of Laurent or $p(z)$-adic expansions which are then plugged into eq.~\eqref{eq:diffeq1} to set up linear systems for the coefficients of $\psi$, which are ultimately used to take the sum of residues contributing to the intersection numbers.  All of these are clearly combinations of rational operations.  The non-trivial part, that we discuss in the following, is how to generate the input for each step of the recursion without reconstructing (potentially large) intermediate results of the previous steps.

For the univariate case, the input is just the analytic expression for $u(\z)$, or equivalently the polynomials $B_j(\z)$ and the generic exponents $\gamma_j$.

For the multivariate case, we start from a numerical implementation of the required list of all $(n-1)$-variate intersection numbers.  These, via straightforward algebraic operations, are combined to obtain the following quantities
\begin{itemize}
    \item the matrix elements $\mathbf{\Omega}_{ij}$ defined in eq.~\eqref{eq:Omega}
    \item the coefficients $\ph_{L,j}$ of the decomposition into left master integrals of $\la \ph_L|_{n-1}$ needed in  the differential equation~\eqref{eq:diffeqn}
    \item the intersection numbers $\la e^{(L)}_j | \ph_R \ra_{n-1}$ appearing in eq.~\eqref{eq:intnumn}.
\end{itemize}
In the following we call the union of these three lists of quantities $\X_n$, which we are thus able to evaluate numerically (over finite fields) as a function of $z_n$ and the other parameters it depends on, including other integration variables $z_k$ with $k>n$.  It is easy to see that $\X_n$ is the input required for the computation of $n$-variate intersection numbers via the recursive algorithm we illustrated in this paper.  More precisely, we need to know the dependence on $z_n$ of each function $f\in \X_n$.  Moreover, we would like to have the list of factors $p(z_n)$ appearing in eq.~\eqref{eq:Pzn}.  In general, we want to avoid to compute the polynomial factors $p(z_n)$ of the denominators of $\X_n$ every time we evaluate intersection numbers.  Knowing the full analytic dependence of the factors $p(z_n)$ (not just on $z_n$ but on every other parameter) avoids the need to perform polynomial factorization at each numerical evaluation and also allows to significantly simplify the reconstruction in $z_n$.  Moreover, as we will now see, the reconstruction of the full analytic dependence of $p(z_n)$ is generally very simple and can be performed in a relatively small number of numerical evaluations.

In order to set up the recursion in our numerical implementation, we follow these three steps
\begin{enumerate}
    \item We first reconstruct every $f\in \X_n$ in the variable $z_n$ modulo a prime number $p$, while every other parameter is set to a random numerical value.  We thus factorize the denominators of the reconstructed functions, modulo $p$, to obtain a semi-numerical list of factors $p(z_n)$.
    \item We proceed to reconstruct the full analytic dependence of the denominator factors $p(z_n)$.  We first identify a ``simple'' subset $\S_n$ of $\X_n$ such that i) the union of the $z_n$-dependent denominator factors in $\S_n$ coincides with the one in $\X_n$ and ii) the degrees of numerators and denominators of the functions in $\S_n$ are as low as possible. The subset $\S_n$ is easily identified from the $z_n$ reconstruction of point 1.  We thus proceed to reconstruct the $z_n$-dependent denominator factors in $\S_n$ as follows.  For each $f\in\S_n$, we make an ansatz of its dependence on $z_n$ of the form
    \begin{equation}
        f(z_n) = \frac{n(z_n)}{d(z_n)}=\frac{\sum_j n_j\, z_n^j}{\sum_j d_j \, z_n^j}
    \end{equation}
    which, again, is easily built from the result of point $1$.  We stress that the coefficients $n_j$ and $d_j$ will have a rational dependence on other parameters, that is implicit in the previous equation.  We also fix the arbitrary normalization of numerator and denominator by setting the lowest degree coefficient of the denominator to be equal to one, i.e.\ $d_{\textrm{min}(j)}=1$.  This choice also has the effect of moving any factor of $f$ that does \emph{not} depend on $z_n$ into the numerator coefficients $n_j$, hence keeping the coefficients $d_j$ very simple.  We can thus write the previous equation as
    \begin{equation}
         \sum_j n_j\, z_n^j - \sum_j d_j \, z_n^j\, f(z_n) = 0,
    \end{equation}
    which evaluated for several values of of $z_n$ yields a linear system that we solve for $n_j$ and $d_j$.  From numerical solutions of this linear systems, obtained with different values of the parameters they depend on, we  reconstruct the full analytic dependence of the denominator coefficients $d_j$ only.  Given their relative simplicity, this usually requires a small number of evaluations and the reconstruction is thus quite efficient.  By factorizing the full analytic form of these denominators in a computer algebra system, we obtain the full list of factors defined in eq.~\eqref{eq:Pzn}.
    \item By combining the results obtained in steps $1$ and $2$ we can easily map each denominator factor (or each suitable product of factors) appearing in the semi-numerical reconstruction of point $1$ into its fully analytic counterpart reconstructed in step $2$.  This allows us to make an ansatz for each $f\in \X_n$ of the form
    \begin{equation}
        f(z_n) = \sum_{\alpha_1,\ldots, \alpha_m,k} c_{\alpha_1,\ldots, \alpha_m,k}\, \frac{z_n^k}{\prod_{j=1}^m p_j(z_n)^{\alpha_j}},
    \end{equation}
    where the unknown coefficients $c_{\alpha_1,\ldots, \alpha_m,k}$ are independent of $z_n$.  By evaluating $\X_n$ for several values of $z_n$ we obtain a linear system of equations that we solve for the unknown coefficients.  This list of numerically-evaluated coefficients, together with the functional form in $z_n$ given by the ansatz, are used as input for the calculation of $n$-variate intersection numbers, which proceeds following subsection~\ref{sec:multivpz}.
\end{enumerate}

\subsection{Master integrals}
\label{sec:mis}

In each step of the recursive algorithm, we need to select a basis and a dual basis of master integrals, which for simplicity was assumed to be known in advance in the discussion above. In ref.~\cite{Lee:2013hzt} it is shown that the number of master integrals can be found as the number of solutions of a system of rational equations.  As observed in~\cite{Frellesvig:2019uqt} the same number can also be computed as the dimension of a polynomial ideal, which in turn can be found via rational operations.  A guess for the list of master integrals can be made from the list of independent monomials in such ideal, see e.g.\ ref.s~\cite{Weinzierl:2020xyy,amslaurea27132} for more details.

As an alternative, we can also find bases of master integrals with a more pragmatic approach that is based on the computation of intersection numbers.  We may start from a list of integrals which form overcomplete bases (i.e.\ a set of spanning vectors that is not minimal)
\begin{equation}
    \mathcal{\tilde E}_R = \left\{|\tilde e^{(R)}_j\ra \right\}, \qquad \mathcal{\tilde E}_L = \left\{\la \tilde e^{(L)}_j |\right\}
\end{equation}
and compute an ``enlarged metric'' $\mathbf{\tilde C}$
\begin{equation}
    \mathbf{\tilde C}_{ij} = \la \tilde e^{(L)}_i | \tilde e^{(R)}_j \ra
\end{equation}
which is obviously not invertible since the basis is overcomplete.  We then column reduce the matrix (or better, a numerical evaluation of it over finite fields) to find a minimal list of independent columns, which corresponds to a basis of independent dual integrals
\begin{equation}
    \mathcal{E}_L = \left\{\la e^{(L)}_j |\right\}_j = \left\{\la \tilde e^{(L)}_j |\right\}_{j\in\textrm{indep.\ columns}} \subset \mathcal{\tilde E}_L.
\end{equation}
The independent columns can thus be row-reduced and the linearly independent rows correspond to a basis of master integrals
\begin{equation}
    \mathcal{E}_R = \left\{|e^{(R)}_j \ra\right\}_j = \left\{|\tilde e^{(R)}_j \ra\right\}_{j\in\textrm{indep.\ rows}} \subset \mathcal{\tilde E}_R.
\end{equation}
In each step of the recursive algorithm, this strategy is used to find a basis $\mathcal{E}_R$ and a dual basis $\mathcal{E}_L$ of master integrals.

 \section{Examples}
\label{sec:ex}
\label{sect:examples}
We present some one- and two-loop examples of reduction to master integrals that use the new rational algorithm presented in this paper.  All the examples we showcase also use our new prescription of section~\ref{sec:rhotrick} for dealing with analytic regulators.  The two methods are however independent of each other and our new rational algorithm has also been extensively tested on several examples retaining the full dependence on analytic regulators (see also~\cite{amslaurea27132}).

In the following subsections we report some of the decompositions we obtained and the bases of master integrals we used, including the ones for intermediate layers of integration.  We also include, in some cases, the reconstructed metric of intersection numbers between master integrals. We stress that one never needs to analytically reconstruct the metric when performing a reduction and we report it in some examples for the sole purpose of showing their simplicity and block triangular structure, a feature that is absent when keeping the full dependence on analytic regulators.

In the following, we use the notation in eq.~\eqref{eq:feynint} to identify Feynman integrals $|\varphi_R\rangle$.  As in the previous sections, the integration variables are $z_1,\ldots,z_n$ and in the recursive algorithm we choose the ordering of integration where $z_1$ is the innermost integration variable and $z_n$ the outermost one.
Even if we do not explicitly write it, to avoid cluttering the notation, we understand that all dual integrals $\la \varphi_L |$ are multiplied by analytic regulators according to the prescription in section~\ref{sec:rhotrick}.  In other words, the replacement
\begin{equation}
    \frac{1}{z_1^{\alpha_1} \dots z_n^{\alpha_n}} \to \rho_1^{\Theta(\alpha_1-\frac{1}{2})}\cdots \rho_n^{\Theta(\alpha_n-\frac{1}{2})}\frac{1}{z_1^{\alpha_1} \dots z_n^{\alpha_n}} \label{eq:rhophilrepl}
\end{equation}
is understood for all the integrands of \emph{left integrals} (i.e.\ \emph{dual integrals}).
For all families we produced reductions for a wide selection of integrals with various combinations of positive and negative powers of denominators.  All of these have been successfully checked against the decomposition obtained using the traditional Laporta algorithm.  A small selection among the simplest ones is quoted for illustration purposes.

\subsection{One-loop families}
\label{subsect:1loop}
\begin{figure}[H]
    \centering
    \includegraphics[scale=0.4]{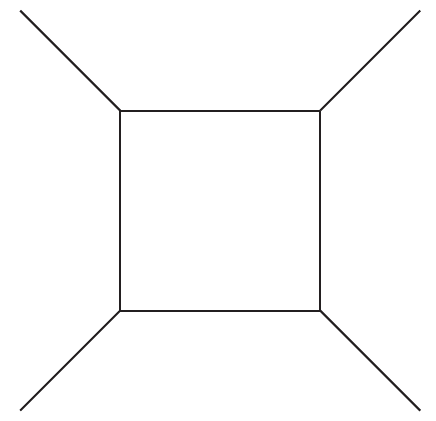}
    \caption{One-loop massless box.}
    \label{fig:box}
\end{figure}

\begin{figure}[H]
    \centering
    \includegraphics[scale=0.4]{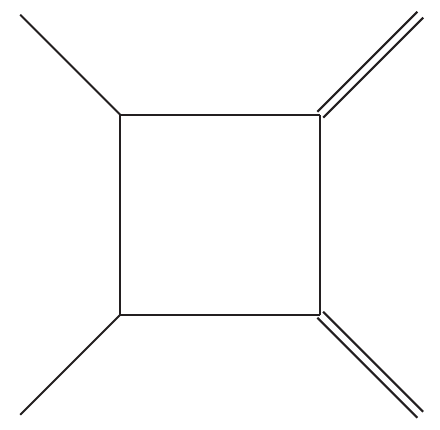}
    \caption{One-loop two-mass hard box.}
    \label{fig:hardbox}
\end{figure}

At one-loop, we consider the integral families of the massless box (Fig.~\ref{fig:box}) and the one-loop box with two off-shell external legs -- also known as \textit{two-mass hard box} (Fig.~\ref{fig:hardbox}). We compute the reduction of several integrals belonging to these two families.

\subsubsection*{Massless box}
We consider the family of one-loop massless box with four external legs $p_1,\ldots, p_4$  satisfying
\begin{equation}
    p_i^2 {}=0 \quad (i=1,\dots,4),\\
\end{equation}
and momentum conservation
\begin{equation}
        p_1+p_2+p_3+p_4 =0.
\end{equation}
The integrals in this family depend on two Mandelstam invariants
\begin{align}
   s  = & ( p_1+p_2)^2, \nonumber \\
   t = & ( p_1+p_3)^2 .
\end{align}
The loop propagators $z_i$ are
\begin{align}
    z_1 &= k_1^2, & z_2&=(k_1-p_1)^2, \nonumber \\
    z_3 &= (k_1-p_1-p_2)^2, & z_4&=(k_1-p_1-p_2-p_3)^2 \label{eq:boxden}
\end{align}
and yield the Baikov polynomial
\begin{align}
    B=\frac{1}{16}\Big(&s^4+2 s^3 t-2 s^3 z_1+2 s^3 z_2-2 s^3 z_3+2 s^3 z_4+s^2 t^2-4 s^2 t z_1+2 s^2 t z_2 \nonumber
    \\
    &-4 s^2 t z_3+2 s^2 t z_4+s^2 z_1^2+s^2 z_2^2+s^2 z_3^2+s^2 z_4^2-2 s^2 z_1 z_2+2 s^2 z_1 z_3 \nonumber\\
    &-2 s^2 z_2 z_3-2 s^2 z_1 z_4+2 s^2 z_2 z_4-2 s^2 z_3 z_4-2 s t^2 z_1-2 s t^2 z_3+2 s t z_1^2 \nonumber\\
    &+2 s t z_3^2-2 s t z_1 z_2-2 s t z_2 z_3-2 s t z_1 z_4+4 s t z_2 z_4-2 s t z_3 z_4+t^2 z_1^2\nonumber\\
    &+t^2 z_3^2-2 t^2 z_1 z_3\Big),
\end{align}

We start with the one-fold integral in $z_1$.  The input for univariate intersection numbers is  the Baikov polynomial and its exponent $\gamma = 1/2 (d-5)$.  The bases for both left and right integrals are given by 
\begin{equation}
    \left\{e^{(L)}_j\right\} = \left\{e^{(R)}_j\right\} = \left\{ \frac{1}{z_1},1 \right\},
\end{equation}
where, as stated above, for left integrals the replacement in eq.~\eqref{eq:rhophilrepl} is understood.
The metric obtained by calculating the univariate intersection numbers between the first-layer master integrals is
\begin{equation}
\mathbf{C}=
\left(
\begin{array}{cc}
 1 & 0 \\
 -\frac{s^2+s (t+z_2-z_3+z_4)+t z_3}{(d-6) (s+t)} & -\frac{4 (d-5) s t (s z_3+t z_3+(z_2-z_3) (z_3-z_4))}{(d-6) (d-4) (s+t)^2} \\
\end{array}
\right).
\end{equation}
Using one-fold intersection numbers, we are able to reduce the one-fold integrals in $z_1$ to master integrals and compute the differential equation matrix w.r.t.\ $z_2$, defined as in eq.~\eqref{eq:Omega}, which reads
\begin{equation}
  \mathbf{\Omega}=  
\left(
\begin{array}{cc}
 -\frac{(d-5) s \left(s^2+s (t+z_2-z_3+z_4)-t (z_3-2 z_4)\right)}{p_2(\mathbf{z})} & 0 \\
 \frac{z_4 (s (s+t-z_2)+t z_3)+z_3 (s (s+t+z_2)-z_3 (s+t))+s z_4^2}{2 (s+t) p_1(\mathbf{z})} & -\frac{(d-6) (z_3-z_4)}{2 p_1(\mathbf{z})} \\
\end{array}
\right),
\end{equation}
where the two $z_2$-dependent polynomial factors in the denominators are
\begin{align}
    p_1(\mathbf{z})=&z_3 (s+t-z_3+z_4)+z_2 (z_3-z_4), \nonumber\\
    p_2(\mathbf{z})=&s^2 z_2^2 + 2 s z_2 (s^2+s t-s z_3+s z_4-t z_3+2 t z_4) +(s^2+s t-s z_3+s z_4-t z_3)^2.
\end{align}
We notice that both the metric and
$\mathbf{\Omega}_{ij}$ present a triangular structure.  This is a general feature of our approach for dealing with analytic regulators.  We also observe that $p_2$ is quadratic in $z_2$, as well as in the other variables $z_i$ it depends on.  Hence, even in this simple example, $\mathbf{\Omega}_{ij}$ has non-rational poles.

After calculating the relevant univariate intersection numbers, we move to the second integration variable, $z_2$, and we compute the needed bivariate intersection numbers. The bases for both left and right integrals for the second layer are given by
\begin{equation}
    \left\{e^{(L)}_j\right\} = \left\{e^{(R)}_j\right\} = \left\{ \frac{1}{z_1 z_2 },\frac{1}{z_1},\frac{1}{z_2} \right\}.
\end{equation}
The metric of bivariate intersection numbers reads
\begin{equation}
\mathbf{C}=
    \left(
\begin{array}{ccc}
 1 & 0 & 0 \\
 \frac{s^2+s (t-z_3+z_4)-t (z_3-2 z_4)}{(d-6) s} & -\frac{4 (d-5) t z_4 (s+t) (s-z_3+z_4)}{(d-6) (d-4) s^2} & 0 \\
 -\frac{s^2+s (t-z_3+z_4)+t z_3}{(d-6) (s+t)} & 0 & -\frac{4 (d-5) s t z_3 (s+t-z_3+z_4)}{(d-6) (d-4) (s+t)^2} \\
\end{array}
\right).
\end{equation}
For the third integration variable $z_3$, the bases for both left and right integrals are
\begin{equation}
    \left\{e^{(L)}_j \right\} = \left\{e^{(R)}_j\right\} = \left\{ \frac{1}{z_1 z_2 z_3 },\frac{1}{z_1 z_3},\frac{1}{z_2} \right\}
\end{equation}
and the metric is
\begin{equation}
\mathbf{C}=
    \left(
\begin{array}{ccc}
 1 & 0 & 0 \\
 \frac{s^2+s (t+z_4)+2 t z_4}{(d-6) s} & -\frac{4 (d-5) t z_4 (s+t) (s+z_4)}{(d-6) (d-4) s^2} & 0 \\
 \frac{s (s+t+z_4)^2}{(d-7) (d-6) (s+t)} & 0 & \frac{s t (s+t+z_4)^4}{4 (d-7) (d-3) (s+t)^2} \\
\end{array}
\right).
\end{equation}

We finally move on to the fourth and last integration variable, $z_4$. The left and right bases for this integral family are
\begin{equation}
    \left\{e^{(L)}_j\right\} = \left\{e^{(R)}_j\right\} = \left\{ \frac{1}{z_1 z_2 z_3 z_4},\frac{1}{z_1 z_3},\frac{1}{z_2 z_4} \right\},
\end{equation}
corresponding to one box and two bubbles. This last step produces the following metric
\begin{equation}
\mathbf{C}=
    \left(
\begin{array}{ccc}
 1 & 0 & 0 \\
 \frac{s (s+t)}{(d-7) (d-6)} & -\frac{s^2 t (s+t)}{4 (d-7) (d-3)} & 0 \\
 \frac{s (s+t)}{(d-7) (d-6)} & 0 & \frac{s t (s+t)^2}{4 (d-7) (d-3)} \\
\end{array}
\right).
\end{equation}
An example of a simple reduction identity is
\begin{align}
    I[1,2,1,2]=&-\frac{4 (d-8) (d-5) (d-3)}{(d-6) s^2 (s+t)^2}  I[1,0,1,0] \nonumber\\
    &-\frac{8 (d-5) (d-3) }{(d-6) s (s+t)^3} I[0,1,0,1] \nonumber \\
    &+\frac{(d-5) ((d-6) s+2 t) }{s (s+t)^2}I[1,1,1,1].
\end{align}
For testing purposes, we also repeated this calculation (and several others) without using the method of analytic regulators described in section~\ref{sec:rhotrick} and retaining the full dependence on the regulators, which we set to be equal to each other, i.e.\ $\rho_j=\rho$ for all $j=1,\ldots,4$.  In this case the metric $\mathbf{C}_{ij}$ and the matrix $\mathbf{\Omega}_{ij}$ are dense and their entries significantly more complicated.  Moreover, at the third layer of integration we have four master integrals instead of just three.  These features are \emph{not} particular to these example and display some of the general advantages of the strategy illustrated in section~\ref{sec:rhotrick}.

\subsubsection*{Two-mass hard box}
We consider the family of the one-loop massless box with two adjacent massive external legs $p_3$ and $p_4$, with $p_3^2 \neq p_4^2$.  This is known in the literature as \emph{two-mass hard box}.  The external momenta $p_1,\dots,p_4$ satisfy
\begin{align}
    p_1^2 {}&=0, \, p_2^2 {}=0 ,\nonumber \\
    p_3^2 {}& \neq 0, \, p_4^2 {} \neq 0, \nonumber \\
     p_3^2 {}& \neq p_4^2,
\end{align}
and momentum conservation
\begin{equation}
        p_1+p_2+p_3+p_4 =0.
\end{equation}
The integrals belonging to this family depend on the external masses $p_3^2$ and $p_4^2$ and the following additional independent invariants
\begin{align}
   s =& ( p_1+p_2)^2, \\
   t =& ( p_1+p_3)^2.
\end{align}
The denominators $z_i$ are defined as in eq.~\eqref{eq:boxden} (but note that now $p_3^2\neq 0$ and $p_4^2\neq 0$).  The Baikov polynomial is
\begin{align}
B=
 \frac{1}{16} &\Bigl(p_3^4 (s+z_2-z_3)^2+2 p_3^2 \bigl(p_4^2 (s^2-s (z_1+z_3)+(z_1-z_2) (z_2-z_3))-s^3\nonumber \\
 &-s^2 (t-z_1+2 z_2-2 z_3+z_4)+s t (z_1-z_2+2 z_3)+s (z_2-z_3) (z_1-z_2+z_3-z_4)\nonumber \\
 &+t (z_1-z_3) (z_3-z_2)\bigr)+p_4^4 (s-z_1+z_2)^2-2 p_4^2 \bigl(s^3+s^2 (t-2 z_1+2 z_2-z_3+z_4)\nonumber \\
 &+s t (-2 z_1+z_2-z_3)+s (z_1-z_2) (z_1-z_2+z_3-z_4)+t (z_1-z_2) (z_1-z_3)\bigr)\nonumber \\
 &+s^4+2 s^3 t-2 s^3 z_1+2 s^3 z_2-2 s^3 z_3+2 s^3 z_4+s^2 t^2-4 s^2 t z_1+2 s^2 t z_2-4 s^2 t z_3\nonumber \\
 &+2 s^2 t z_4+s^2 z_1^2+s^2 z_2^2+s^2 z_3^2+s^2 z_4^2-2 s^2 z_1 z_2+2 s^2 z_1 z_3-2 s^2 z_2 z_3-2 s^2 z_1 z_4\nonumber \\
 &+2 s^2 z_2 z_4-2 s^2 z_3 z_4-2 s t^2 z_1-2 s t^2 z_3+2 s t z_1^2+2 s t z_3^2-2 s t z_1 z_2-2 s t z_2 z_3\nonumber \\
 &-2 s t z_1 z_4+4 s t z_2 z_4-2 s t z_3 z_4+t^2 z_1^2+t^2 z_3^2-2 t^2 z_1 z_3\Bigr),
\end{align}
There are four integration variables $z_1,\ldots,z_4$.  The bases of $k$-fold master integrals for each layer $k$ of integration are given by
\begin{align}
1-\text{fold} \nonumber\\ 
    &\left\{e^{(L)}_j\right\} = \left\{e^{(R)}_j\right\} = \left\{ \frac{1}{z_1},1 \right\}, \displaybreak[0] \nonumber \\
2-\text{fold} \nonumber\\ 
    &\left\{e^{(L)}_j\right\} = \left\{e^{(R)}_j\right\} = \left\{ \frac{1}{z_1 z_2 },\frac{1}{z_2},\frac{1}{z_1},1 \right\}, \displaybreak[0] \nonumber \\
3-\text{fold} \nonumber\\ 
    &\left\{e^{(L)}_j \right\} = \left\{e^{(R)}_j\right\} = \left\{ \frac{1}{z_1 z_2 z_3 },\frac{1}{z_1 z_3},\frac{1}{z_3},\frac{1}{z_2},\frac{1}{z_1},1 \right\}, \displaybreak[0] \nonumber \\
4-\text{fold} \nonumber\\ 
    &\left\{e^{(L)}_j\right\} = \left\{e^{(R)}_j\right\} = \left\{ \frac{1}{z_1 z_2 z_3 z_4},\frac{1}{z_1 z_3 z_4},\frac{1}{z_3 z_4},\frac{1}{z_2 z_4},\frac{1}{z_1 z_4},\frac{1}{z_1 z_3}\right\}.
\end{align}
The master integrals for the last layer coincide with the one of the integral family and consist of one box, one triangle and four bubbles.  The metric presents the following triangular structure
\begin{align}
\mathbf{C}=
\left(
\begin{array}{cccccc}
 C_{11} & 0 & 0 & 0 & 0 & 0 \\
 C_{21}& C_{22} & 0 & 0 & 0 & 0 \\
 C_{31} & C_{32} &C_{33} & 0 & 0 & 0 \\
 C_{41} & 0 & 0 & C_{44} & 0 & 0 \\
 C_{51} & C_{52} & 0 & 0 &C_{55} & 0 \\
 C_{61} & C_{62} & 0 & 0 & 0 & C_{66} \\
\end{array}
\right),
\end{align}
where the $C_{ij}$ are rational functions of the kinematic invariants and of the dimensional regulator.  We report a simple representative reduction for this family
\begin{align}
    I[1,1,1,2]=&\frac{(d-4) \left(p_3^4-2 p_3^2 (p_4^2+s)+(p_4^2-s)^2\right) I[1,0,1,1]}{2 p_3^2 p_4^2 s (p_3^2+p_4^2-s-t)} \nonumber \\
    &+\frac{(d-3) (p_3^2-p_4^2-s) }{p_3^2 p_4^2 s (p_3^2+p_4^2-s-t)}I[0,0,1,1] \nonumber \\
    &+\frac{(d-3) (p_3^2-p_4^2+s)}{p_3^2 p_4^2 s (-p_3^2-p_4^2+s+t)} I[1,0,0,1] \nonumber \\
    &+\frac{(d-3) (p_3^2+p_4^2-s)}{p_3^2 p_4^2 s (-p_3^2-p_4^2+s+t)} I[1,0,1,0] \nonumber \\
    &+\frac{(d-5)}{-p_3^2-p_4^2+s+t} I[1,1,1,1].
\end{align}

\subsection{Two-loop families}
\begin{figure}[H]
    \centering
    \includegraphics[scale=0.6]{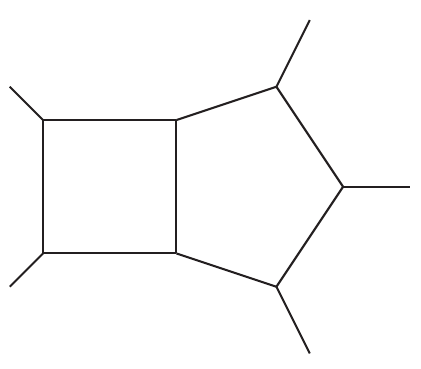}
    \caption{Massless pentabox, in the example we consider its maximal cut.}
    \label{fig:pentabox}
\end{figure}

We consider the decomposition of integrals belonging to the families of two-loop massless kite (Fig.~\ref{fig:kite}), the two-loop massive sunrise with equal internal masses (Fig.~\ref{fig:sun}) and the two-loop massless pentabox on the maximal cut (Fig.~\ref{fig:pentabox}).
\label{subsect:2loop}

\subsubsection*{Massless two-loop kite}
\label{subsubsect:2bubble}
\begin{figure}[H]
    \centering
    \includegraphics[scale=0.5]{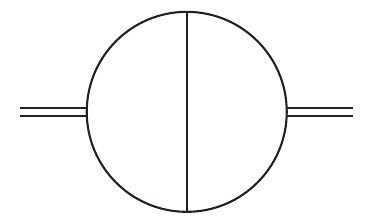}
    \caption{Two loop massless kite with off-shell external legs.}
    \label{fig:kite}
\end{figure}
We consider the family of two-loop massless kite (Fig.~\ref{fig:kite}) with off-shell external momentum~$p$.
The integrals depend on the invariant
\begin{align}
   s=p^2.
\end{align}
The propagators are
\begin{align}
    z_1 &= k_1^2, & z_2&=(k_1-p_1)^2, & z_3=(k_1-k_2)^2, \nonumber \\
    z_4&=k_2^2, & z_5&=(k_2-p_1)^2
\end{align}
and the Baikov polynomial is given by
\begin{align}
    B=\frac{1}{4} \Big(&-s^2 z_3-s z_3^2+s z_1 z_3+s z_2 z_3+s z_2 z_4+s z_3 z_4+s z_1 z_5+s z_3 z_5-s z_4 z_5 \nonumber\\
    &-s z_1 z_2-z_2 z_4^2-z_1 z_5^2-z_2^2 z_4+z_1 z_2 z_4-z_1 z_3 z_4+z_2 z_3 z_4-z_1^2 z_5+z_1 z_2 z_5 \nonumber\\
    &+z_1 z_3 z_5-z_2 z_3 z_5+z_1 z_4 z_5+z_2 z_4 z_5\Big)
\end{align}
We present the basis of $k$-fold master integrals for each layer $k$ of integration, highlighting that in some cases the left and right bases may not be the same
\begin{align}
    1-\text{fold} \nonumber\\ &\left\{e^{(L)}_j\right\} = \left\{e^{(R)}_j\right\} = \left\{ 1, \frac{1}{z_1}\right\}, \nonumber \displaybreak[0] \\
    2-\text{fold} \nonumber \\ &\left\{e^{(L)}_j\right\} = \left\{e^{(R)}_j\right\} = \left\{ \frac{1}{z_1 z_2}, \frac{1}{z_1}, \frac{1}{z_2}, 1\right\},
    \nonumber \displaybreak[0] \\
    3-\text{fold} \nonumber\\ &\left\{e^{(L)}_j\right\} = \left\{e^{(R)}_j\right\} = \left\{ \frac{1}{z_1 z_2 z_3}, \frac{1}{z_2 z_3}, \frac{1}{z_1 z_3}, \frac{1}{z_1 z_2}\right\},
    \nonumber \displaybreak[0] \\
    4-\text{fold} \nonumber\\ &\left\{e^{(L)}_j\right\} = \left\{ \frac{1}{z_2 z_3 z_4},\frac{1}{z_1 z_2 z_4}, \frac{1}{z_2 z_3}, \frac{1}{z_1 z_3}, \frac{1}{z_1 z_2}, \frac{1}{z_3} \right\} \nonumber\\
    &\left\{e^{(R)}_j\right\} = \left\{ \frac{1}{z_2 z_3 z_4},\frac{1}{z_1 z_2 z_4}, \frac{1}{z_1 z_2 z_3}, \frac{1}{z_2 z_3}, \frac{1}{z_1 z_3}, \frac{1}{z_1 z_2} \right\},
    \nonumber \displaybreak[0] \\
    5-\text{fold} \nonumber\\ &\left\{e^{(L)}_j\right\} = \left\{e^{(R)}_j\right\} = \left\{ \frac{1}{z_1 z_2 z_4 z_5 }, \frac{1}{z_1 z_3 z_5 },  \frac{1}{z_2 z_3 z_4 }\right\}.
\end{align}
The metric for the last layer of master integrals reads
\begin{equation}
\mathbf{C}=
    \left(
\begin{array}{ccc}
 -\frac{(d-4) s^2}{4 (d-5) (d-3)} & 0 & 0 \\
 0 & \frac{4 (d-4)^2 s^4}{3 (3 d-16) (3 d-14) (3 d-10) (3 d-8)} & 0 \\
 0 & 0 & \frac{4 (d-4)^2 s^4}{3 (3 d-16) (3 d-14) (3 d-10) (3 d-8)} \\
\end{array}
\right)
\end{equation}
and presents a diagonal structure due to the fact that no master integral belongs to a subsector of the others.  We quote a simple example of a reduction for this family
\begin{align}
    I[1,1,2,1,1]={}&-\frac{3 (d-5) (d-2) (3 d-10) (3 d-8)}{(d-6)^2 (d-4) s^3} I[0,1,1,1,0] \nonumber \\
    &-\frac{3 (d-5) (d-2) (3 d-10) (3 d-8)}{(d-6)^2 (d-4) s^3} I[1,0,1,0,1] \nonumber \\
    &+\frac{4 (d-3) }{(d-6) s^2}I[1,1,0,1,1].
\end{align}

\subsubsection*{Massive two-loop sunrise}
\begin{figure}[H]
    \centering
    \includegraphics[scale=0.5]{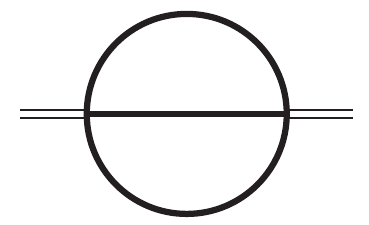}
    \caption{Two loop massive sunrise with equal internal masses and off-shell external momenta.}
    \label{fig:sun}
\end{figure}
We consider the family of two-loop massive sunrise (Fig.~\ref{fig:sun}) with equal-mass internal lines and off-shell external momentum $p$.
The integrals depend on the internal mass $m$ and on the external invariant
\begin{align}
    s = p^2.
\end{align}
The propagators are
\begin{align}
    z_1&=(k_1+p_1)^2, & z_2&=k_1^2-m^2, & z_3&=(k_1+k_2+p_1)^2-m^2, \nonumber \\
    z_4&=(k_2+p_1)^2, & z_5&=k_2^2-m^2
\end{align}
and the Baikov polynomial is given by
\begin{align}
    B=\frac{1}{4}\Big(&-m^6+2 m^4 s-2 m^4 z_2+m^4 z_3-2 m^4 z_5-m^2 s^2+2 m^2 s z_2+2 m^2 s z_5 \nonumber\\
    &-m^2 z_2^2-m^2 z_5^2+m^2 z_1 z_2-m^2 z_1 z_3+m^2 z_2 z_3+3 m^2 z_1 z_4-m^2 z_3 z_4 \nonumber\\
    &-3 m^2 z_2 z_5+m^2 z_3 z_5+m^2 z_4 z_5-s^2 z_3-s z_3^2-s z_1 z_2+s z_1 z_3+s z_2 z_3 \nonumber \\
    &+s z_1 z_4+s z_3 z_4+s z_2 z_5+s z_3 z_5-s z_4 z_5-z_1 z_4^2-z_2 z_5^2-z_1^2 z_4+z_1 z_2 z_4 \nonumber \\
    &+z_1 z_3 z_4-z_2 z_3 z_4-z_2^2 z_5+z_1 z_2 z_5-z_1 z_3 z_5+z_2 z_3 z_5+z_1 z_4 z_5+z_2 z_4 z_5\Big).
\end{align}
We consider integrals of the form in~\eqref{eq:feynint} with the constrain
\begin{equation}
    \alpha_1 \leq 0, \qquad \alpha_4 \leq 0
\end{equation}
i.e.\ we identify $z_1$ and $z_4$ as auxiliary denominators.
We report the bases for each layer of integration, highlighting that in some cases the basis for left and right integrals may not be the same.
\begin{align}
    1-\text{fold} \nonumber\\ &\left\{e^{(L,1)}_j\right\} = \left\{e^{(R,1)}_j\right\} = \left\{ 1 \right\}, \displaybreak[0] \\
    2-\text{fold} \nonumber\\ &\left\{e^{(L,2)}_j\right\} = \left\{e^{(R,2)}_j\right\} = \left\{ \frac{1}{z_2}, 1\right\}, \displaybreak[0]
    \\
    3-\text{fold} \nonumber\\ &\left\{e^{(L,3)}_j\right\} = \left\{e^{(R,3)}_j\right\} = \left\{ \frac{1}{z_2 z_3}, \frac{1}{z_3}, \frac{1}{z_2}\right\},
    \displaybreak[0] \\
    4-\text{fold} \nonumber\\ &\left\{e^{(L,4)}_j\right\} = \left\{ \frac{1}{z_2^2},\frac{1}{z_2 z_3}, \frac{1}{z_3}, \frac{1}{z_2}, z_3, 1\right\} \nonumber \\
    &\left\{e^{(R,4)}_j\right\} = \left\{ \frac{1}{z_2 z_3^2},\frac{1}{z_2^2 z_3}, \frac{z_4}{ z_2 z_3},\frac{1}{z_2 z_3}, \frac{1}{z_3}, \frac{1}{z_2} \right\},
    \displaybreak[0] \\
    5-\text{fold} \nonumber\\ &\left\{e^{(L,5)}_j\right\} = \left\{e^{(R,5)}_j\right\} = \left\{ \frac{1}{z_2 z_3 z_5^2 }, \frac{1}{z_2^2 z_3 z_5 },\frac{1}{z_2 z_3 z_5 },  \frac{1}{z_3 z_5},\frac{1}{z_2 z_5},\frac{1}{z_2 z_3}\right\}.
\end{align}
The $5$-fold metric presents the following block triangular structure
\begin{align}
\mathbf{C}=
\left(
\begin{array}{cccccc}
 C_{11} & C_{12} & C_{13} & 0 & 0 & 0 \\
 C_{21}& C_{22} & C_{23} & 0 & 0 & 0 \\
 C_{31} & C_{32} &C_{33} & 0 & 0 & 0 \\
 C_{41} & C_{42} & C_{43} & C_{44} & 0 & 0 \\
 C_{51} & C_{52} & C_{53} & 0 &C_{55} & 0 \\
 C_{61} & C_{62} & C_{63} & 0 & 0 & C_{66} \\
\end{array}
\right)
\end{align}
Here is the example of a reduction
\begin{align}
  I[0,1,1,-1,1]=&\frac{1}{48 (d-5) (d-3) m^4 s^2}\Big(-\left((d-4) (d-2) m^8\right)+2 (d-2) (3 d-14) m^6 s \nonumber\\
  &+16 (d-5) (d-3) m^4 s^2 -2 (d-2) (3 d-14) m^2 s^3 \nonumber \\
  & +(d-4) (d-2) s^4 \Big)I[0,0,1,0,1]  \nonumber \displaybreak[0] \\
  &+\frac{1}{48 (d-5) (d-3) m^4 s^2}\Big((d-4) (d-2) m^8-2 (d-2) (3 d-14) m^6 s \nonumber\\
  &+16 (d-5) (d-3) m^4 s^2+2 (d-2) (3 d-14) m^2 s^3 \nonumber \\
  & -(d-4) (d-2) s^4 \Big)I[0,1,0,0,1] \nonumber \displaybreak[0] \\
  &+\frac{4 m^2 (s-m^2) I[0,1,1,0,2]}{3 (d-2)}+\frac{4 m^2 (m^2-s) I[0,2,1,0,1]}{3 (d-2)} \nonumber\\
  &+\left(m^2+\frac{s}{3}\right) I[0,1,1,0,1]+\frac{1}{3} I[0,1,1,0,0].
\end{align}
We observe that the master integrals in this example obey additional symmetry relations which cannot be cast as IBP identities.  Indeed, as already stated in section~\ref{sec:int-th}, the framework of intersection theory does not take these into account.  This is not an issue, because these additional relations can be easily identified and added afterwards.  In this case, one can easily see that integrals of the form $I[0,\alpha_2,\alpha_3,0,\alpha_5]$
are invariant under permutations of the exponents $\alpha_j$.  This reduces the number of independent integrals from six to three, since
\begin{equation}
    I[0, 1, 1, 0, 2] = I[0, 2, 1, 0, 1], \qquad
    I[0, 1, 1, 0, 0] = I[0, 1, 0, 0, 1] = I[0,0,1,0,1].
\end{equation}

\subsubsection*{Massless pentabox on the maximal cut}
We consider the family of massless pentabox on the maximal cut. The external legs $p_1,\dots,p_5$ satisfy
\begin{align}
    p_i^2 {}=0,
\end{align}
and momentum conservation
\begin{equation}
        p_1+p_2+p_3+p_4+p_5=0.
\end{equation}
The integrals belonging to this family depend on five independent invariants which we choose as
\begin{equation}
    \left\{ s_{12}, s_{23}, s_{34}, s_{45}, s_{51} \right\},
\end{equation}
where
\begin{equation}
    s_{ij} = (p_i+p_j)^2.
\end{equation}
The pentabox family depends on $11$ generalized denominators
\begin{align}
    z_1&=(k_2+p_1+p_2)^2, & z_2&=(k_2+p_1)^2, & z_3&=(k_1+p_5)^2, \nonumber \\
    z_4&=k_1^2, & z_5&=(k_1-p_1)^2, & z_6&=(k_1-p_1-p_2)^2, \nonumber \\
    z_7&=(k_1-p_1-p_2-p_3)^2,& z_8&=k_2^2, & z_9&=(k_2+p_1+p_2+p_3+p_4)^2, \nonumber \\
    z_{10}&=(k_2+p_1+p_2+p_3)^2, & z_{11}&=(k_1+k_2)^2.
\end{align}
of which the first three are irreducible scalar products, i.e.\ we consider integrals of the form~\eqref{eq:feynint} with
\begin{equation}
    \alpha_1\leq 0, \quad \alpha_2\leq 0, \quad \alpha_3\leq 0.
\end{equation}
We consider the reduction of integrals belonging to the maximal cut of this family, which corresponds to putting on-shell all the denominators via the substitution
\begin{align}
    \frac{1}{z_i} \rightarrow \delta(z_i) \quad (i=4,\dots,11)
\end{align}
in the expressions for $\ph_{R,L}$.  This still yields the correct coefficients of reductions multiplying the master integrals of the top sector.
Therefore the integrals we reduce are functions only of the auxiliary denominators $z_1,z_2,z_3$, meaning that we calculate intersection numbers of three-fold integrals.
The Baikov polynomial of pentabox on the maximal-cut is given by
\begin{align}
B = \frac{1}{64} &\Big(-z_3 (s_{12}^2 z_3 (s_{23}-z_2)^2+2 s_{12} (-s_{23}^2 z_1 z_3+s_{23} (s_{45} z_2 z_3+s_{45} z_1 (2 z_2+z_3) \nonumber \\
&+z_1 z_2 z_3)+s_{45} (z_1-z_2) z_2 z_3)+z_3 (s_{23} z_1+s_{45} (z_2-z_1))^2) \nonumber \\
&+2 s_{15} s_{45} z_1 (z_3 (s_{12} (s_{23}+z_2)-s_{23} z_1+s_{45} z_1-s_{45} z_2)+s_{34} s_{45} z_2) \nonumber \\
&+2 s_{34} s_{45} z_2 z_3 (s_{12} (s_{23}-z_2)+s_{23} z_1+s_{45} (z_2-z_1))-s_{15}^2 s_{45}^2 z_1^2-s_{34}^2 s_{45}^2 z_2^2\Big),
\end{align}
and its exponent is $\gamma = 1/2 (d-7)$.  The bases of master integrals for each layer of integration are
\begin{align}
1-\text{fold} \nonumber\\ 
    &\left\{e^{(L)}_j\right\} = \left\{e^{(R)}_j\right\} = \left\{  1 \right\}, \nonumber \\
2-\text{fold} \nonumber\\ 
    &\left\{e^{(L)}_j\right\} = \left\{e^{(R)}_j\right\} = \left\{ 1 \right\}, \nonumber \\
3-\text{fold} \nonumber\\ 
    &\left\{e^{(L)}_j \right\} = \left\{e^{(R)}_j\right\} = \left\{ z_2, z_3, 1\right\}.
\end{align}
Since we are on the maximal cut, all the integrals we are considering belong to the same sector and thus all entries of the metric $\mathbf{C}_{ij}$ are non-vanishing.  We reconstructed the reduction of all integrals up to degree 5 in the variables $z_i$ (which are those that contribute e.g.\ to QCD amplitudes) and successfully checked them against the Laporta method.

\section{Conclusions and Outlook}
\label{sec:conclusions}

We presented a new method for computing intersection numbers via a purely rational algorithm that does not require any change of basis or integral transformation.  This is achieved via the systematic use of the $p(z)$-adic series expansion.  The latter expands functions in powers of a polynomial, allowing to study their behaviour close to potentially irrational singular points, without having to perform any irrational operation or knowing the explicit location of these points.

This result represents significant progress in the application of intersection theory to the decomposition of Feynman integrals.  The new algorithm is elegant and satisfactory from a mathematical and theoretical point of view, since everything is rational in all steps of the calculation.  It also sidesteps the algebraic bottleneck of dealing with irrational expressions and opens up the possibility of combining intersection theory with efficient computational techniques, such as finite fields and functional reconstruction.  As a proof of concept, we implemented the new algorithm over finite fields, using the \textsc{Mathematica} package \textsc{FiniteFlow} and tested it on several one- and two-loop examples.

We also proposed a new strategy for dealing with analytic regulators, de facto removing any analytic dependence of the calculation on them and showing dramatic simplifications on both analytic expressions and the structure of the results.

These results open up several possible lines of research aimed at making intersection theory a viable approach for phenomenological applications.  Besides optimizing and simplifying our proof-of-concept implementation, exploring different integral representations which reduce the number of integration variables may lead to substantial improvements.  Examples are the loop-by-loop Baikov representation and the Lee-Pomeransky representation~\cite{Lee:2013hzt}.  Even more interestingly, a method for the direct calculation of multivariate intersection numbers  -- thus stepping away from the recursive univariate method --  has recently been proposed~\cite{Chestnov:2022xsy}. A possible generalization of our method that builds on~\cite{Chestnov:2022xsy} and a multivariate generalization of the $p(z)$-adic expansion, combined with the multivariate global residue theorem, is worth investigating in the near future.

Finally, we expect the method of $p(z)$-adic expansions may also find uses in a broader spectrum of applications, where one needs to study the behaviour of certain functions close to the zeroes of certain polynomials. For instance, they might be employed in studies of the singular structure of amplitudes for a more efficient reconstruction of them -- as recently done using the similar concept of $p$-adic numbers~\cite{DeLaurentis:2022otd}.  The possibility of systematically employing $p(z)$-adic expansions using publicly available finite-field frameworks~\cite{Peraro:2019svx} and the fact that these are not affected by numerical instabilities, makes this technique a suitable candidate for future theoretical and phenomenological investigations.

\appendix
\section{$p(z)$-adic expansions via polynomial division}
\label{sec:polydiv}

We illustrate how we can obtain the coefficients $c_i(z)$ of the $p(z)$-adic expansion in~\eqref{eq:pzadic} as reminders of polynomial divisions. Since $f(z)$ is a rational function, we can write it as
\begin{equation}
    f(z) = \frac{n(z)}{p(z)^k\, d(z)},
\end{equation}
where $d(z)$ is co-prime (i.e.\ has no common factor) with $p(z)$. Since the factor $1/p(z)^k$ can be trivially added back after the expansion, it is sufficient to consider a function of the form
\begin{equation}
    f^{\star}(z) = \frac{n(z)}{d(z)},
\end{equation}
whose denominator $d(z)$ is co-prime with $p(z)$.

Let $\tilde d (z)$ be the multiplicative inverse of $d(z)$ with respect to $p(z)$, namely the polynomial of degree lower than $\deg p$ such that
\begin{equation}
    d(z)\, \tilde d(z) = 1 \mod p(z).\label{eq:mulinv}
\end{equation}
One can show that, if $d(z)$ and $p(z)$ are co-prime, the multiplicative inverse of $d(z)$ exists and is unique.  It can be computed using the extended Euclidean algorithm, or simply by making an ansatz for it and fixing the unknown coefficients using eq.~\eqref{eq:mulinv}.

The first term of the $p(z)$-adic expansion of $f^{\star}(z)$ is simply its polynomial reminder modulo $p(z)$
\begin{equation}
    c_0(z) \equiv f^{\star}(z) \, \text{mod} \, p(z) = n(z)\tilde{d}(z) \, \text{mod} \, p(z).
\label{app:fmodp}
\end{equation}
We can thus rewrite the numerator $n(z)$ as
\begin{equation}
    n(z) = c_0(z) d(z) + q_1(z) p(z),
\end{equation}
for some polynomial $q_1(z)$, hence
\begin{equation}
    f^{\star}(z) = c_0(z) + \frac{q_1(z)}{d(z)}\,  p(z).
\end{equation}
To find the next term in the expansion we repeat the procedure considering $q_1(z)/d(z)$ as a new rational function.  In general we have the recursive formula
\begin{equation}
    c_i(z) = \frac{q_i(z)}{d(z)} \, \text{mod}\,p(z),
\end{equation}
where $q_{i}(z)$ for $i>0$ can be computed from $q_{i-1}(z)$ and $c_{i-1}(z)$ using
\begin{equation}
    q_{i-1}(z) = c_{i-1}(z) d(z) + q_{i}(z) p(z),
\end{equation}
and $q_0(z)=n(z)$. Following this procedure we can compute $c_i(z)$ up to the order we need.

Readers familiar with the concept of $p$-adic numbers may recognize that the construction of the $p(z)$-adic expansion of a rational function is analogous to the construction of the $p$-adic expansion of a rational number, where $p$ is a prime number rather than a prime polynomial over the rational field and the remainder of integer division is used instead of the polynomial remainder.

\section{The univariate global residue theorem}
\label{sec:globres}
We give a brief overview of the univariate global residue theorem, whose generalization is a key ingredient of the calculation of intersection numbers with the rational algorithm, as described in subsection~\ref{sec:pzadic}.
Given a rational function 
\begin{equation}
    f(z) = \frac{n(z)}{d(z)},
\end{equation}
whose denominator is coprime with a polynomial $p(z)$, we can calculate the polynomial reminder of $f(z)$ modulus $p(z)$ with eq.~\eqref{app:fmodp}. It takes the explicit form
\begin{equation}
    \tilde{f}(z) \equiv f(z)\, \, \text{mod}\, \, p(z) = \sum_{j=1}^{\text{deg}\, p -1} f_j z^j.
\end{equation}
The univariate global residue theorem states that the global residue of $f(z)/p(z)$, namely the sum of all the local residues of $f(z)/p(z)$ at the zeroes of $p(z)$, is 
\begin{equation}
    \text{Res}_{p(z)} \left( \frac{f(z)}{p(z)} \right) = \frac{f_{\text{deg} \,p -1}}{l_c},
\end{equation}
where $\text{Res}_{p(z)}$ is defined in eq.~\eqref{eq:respzdef}.

\acknowledgments
We thank Vsevolod Chestnov, Federico Gasparotto and Pierpaolo Mastrolia for many valuable discussions and comments on this work.  We also grateful to Thomas Gehrmann and Petr Jakubčík for feedback on the draft.
This work was supported by the European Research Council (ERC) under the European Union’s Horizon $2020$ research and innovation programme grant agreements $101019620$ (ERC Advanced Grant TOPUP) and 101040760 (ERC Starting Grant FFHiggsTop).

\bibliographystyle{JHEP}
\bibliography{biblio}

\providecommand{\href}[2]{#2}\begingroup\raggedright\begin{thebibliography}{10}

\bibitem{Tkachov:1981wb}
F.~V. Tkachov, {\it {A Theorem on Analytical Calculability of Four Loop Renormalization Group Functions}},  {\em Phys. Lett. B} {\bf 100} (1981) 65--68.

\bibitem{Chetyrkin:1981qh}
K.~G. Chetyrkin and F.~V. Tkachov, {\it {Integration by Parts: The Algorithm to Calculate beta Functions in 4 Loops}},  {\em Nucl. Phys. B} {\bf 192} (1981) 159--204.

\bibitem{Gehrmann:1999as}
T.~Gehrmann and E.~Remiddi, {\it {Differential equations for two loop four point functions}},  {\em Nucl. Phys. B} {\bf 580} (2000) 485--518, [\href{http://arxiv.org/abs/hep-ph/9912329}{{\tt hep-ph/9912329}}].

\bibitem{Laporta:2000dsw}
S.~Laporta, {\it {High precision calculation of multiloop Feynman integrals by difference equations}},  {\em Int. J. Mod. Phys. A} {\bf 15} (2000) 5087--5159, [\href{http://arxiv.org/abs/hep-ph/0102033}{{\tt hep-ph/0102033}}].

\bibitem{vonManteuffel:2012np}
A.~von Manteuffel and C.~Studerus, {\it {Reduze 2 - Distributed Feynman Integral Reduction}},  \href{http://arxiv.org/abs/1201.4330}{{\tt arXiv:1201.4330}}.

\bibitem{Smirnov:2008iw}
A.~V. Smirnov, {\it {Algorithm FIRE -- Feynman Integral REduction}},  {\em JHEP} {\bf 10} (2008) 107, [\href{http://arxiv.org/abs/0807.3243}{{\tt arXiv:0807.3243}}].

\bibitem{Maierhofer:2017gsa}
P.~Maierh\"ofer, J.~Usovitsch, and P.~Uwer, {\it {Kira\textemdash{}A Feynman integral reduction program}},  {\em Comput. Phys. Commun.} {\bf 230} (2018) 99--112, [\href{http://arxiv.org/abs/1705.05610}{{\tt arXiv:1705.05610}}].

\bibitem{vonManteuffel:2014ixa}
A.~von Manteuffel and R.~M. Schabinger, {\it {A novel approach to integration by parts reduction}},  {\em Phys. Lett. B} {\bf 744} (2015) 101--104, [\href{http://arxiv.org/abs/1406.4513}{{\tt arXiv:1406.4513}}].

\bibitem{Peraro:2016wsq}
T.~Peraro, {\it {Scattering amplitudes over finite fields and multivariate functional reconstruction}},  {\em JHEP} {\bf 12} (2016) 030, [\href{http://arxiv.org/abs/1608.01902}{{\tt arXiv:1608.01902}}].

\bibitem{Klappert:2019emp}
J.~Klappert and F.~Lange, {\it {Reconstructing rational functions with FireFly}},  {\em Comput. Phys. Commun.} {\bf 247} (2020) 106951, [\href{http://arxiv.org/abs/1904.00009}{{\tt arXiv:1904.00009}}].

\bibitem{Smirnov:2019qkx}
A.~V. Smirnov and F.~S. Chuharev, {\it {FIRE6: Feynman Integral REduction with Modular Arithmetic}},  {\em Comput. Phys. Commun.} {\bf 247} (2020) 106877, [\href{http://arxiv.org/abs/1901.07808}{{\tt arXiv:1901.07808}}].

\bibitem{Peraro:2019svx}
T.~Peraro, {\it {FiniteFlow: multivariate functional reconstruction using finite fields and dataflow graphs}},  {\em JHEP} {\bf 07} (2019) 031, [\href{http://arxiv.org/abs/1905.08019}{{\tt arXiv:1905.08019}}].

\bibitem{Klappert:2020nbg}
J.~Klappert, F.~Lange, P.~Maierh\"ofer, and J.~Usovitsch, {\it {Integral reduction with Kira 2.0 and finite field methods}},  {\em Comput. Phys. Commun.} {\bf 266} (2021) 108024, [\href{http://arxiv.org/abs/2008.06494}{{\tt arXiv:2008.06494}}].

\bibitem{Smirnov:2010hn}
A.~V. Smirnov and A.~V. Petukhov, {\it {The Number of Master Integrals is Finite}},  {\em Lett. Math. Phys.} {\bf 97} (2011) 37--44, [\href{http://arxiv.org/abs/1004.4199}{{\tt arXiv:1004.4199}}].

\bibitem{Mizera:2017rqa}
S.~Mizera, {\it {Scattering Amplitudes from Intersection Theory}},  {\em Phys. Rev. Lett.} {\bf 120} (2018), no.~14 141602, [\href{http://arxiv.org/abs/1711.00469}{{\tt arXiv:1711.00469}}].

\bibitem{Mastrolia:2018uzb}
P.~Mastrolia and S.~Mizera, {\it {Feynman Integrals and Intersection Theory}},  {\em JHEP} {\bf 02} (2019) 139, [\href{http://arxiv.org/abs/1810.03818}{{\tt arXiv:1810.03818}}].

\bibitem{Frellesvig:2019kgj}
H.~Frellesvig, F.~Gasparotto, S.~Laporta, M.~K. Mandal, P.~Mastrolia, L.~Mattiazzi, and S.~Mizera, {\it {Decomposition of Feynman Integrals on the Maximal Cut by Intersection Numbers}},  {\em JHEP} {\bf 05} (2019) 153, [\href{http://arxiv.org/abs/1901.11510}{{\tt arXiv:1901.11510}}].

\bibitem{Frellesvig:2019uqt}
H.~Frellesvig, F.~Gasparotto, M.~K. Mandal, P.~Mastrolia, L.~Mattiazzi, and S.~Mizera, {\it {Vector Space of Feynman Integrals and Multivariate Intersection Numbers}},  {\em Phys. Rev. Lett.} {\bf 123} (2019), no.~20 201602, [\href{http://arxiv.org/abs/1907.02000}{{\tt arXiv:1907.02000}}].

\bibitem{Mizera:2019gea}
S.~Mizera, {\em {Aspects of Scattering Amplitudes and Moduli Space Localization}}.
\newblock PhD thesis, Princeton, Inst. Advanced Study, 2020.
\newblock \href{http://arxiv.org/abs/1906.02099}{{\tt arXiv:1906.02099}}.

\bibitem{Mizera:2019vvs}
S.~Mizera and A.~Pokraka, {\it {From Infinity to Four Dimensions: Higher Residue Pairings and Feynman Integrals}},  {\em JHEP} {\bf 02} (2020) 159, [\href{http://arxiv.org/abs/1910.11852}{{\tt arXiv:1910.11852}}].

\bibitem{Frellesvig:2020qot}
H.~Frellesvig, F.~Gasparotto, S.~Laporta, M.~K. Mandal, P.~Mastrolia, L.~Mattiazzi, and S.~Mizera, {\it {Decomposition of Feynman Integrals by Multivariate Intersection Numbers}},  {\em JHEP} {\bf 03} (2021) 027, [\href{http://arxiv.org/abs/2008.04823}{{\tt arXiv:2008.04823}}].

\bibitem{Weinzierl:2020xyy}
S.~Weinzierl, {\it {On the computation of intersection numbers for twisted cocycles}},  {\em J. Math. Phys.} {\bf 62} (2021), no.~7 072301, [\href{http://arxiv.org/abs/2002.01930}{{\tt arXiv:2002.01930}}].

\bibitem{Chestnov:2022alh}
V.~Chestnov, F.~Gasparotto, M.~K. Mandal, P.~Mastrolia, S.~J. Matsubara-Heo, H.~J. Munch, and N.~Takayama, {\it {Macaulay matrix for Feynman integrals: linear relations and intersection numbers}},  {\em JHEP} {\bf 09} (2022) 187, [\href{http://arxiv.org/abs/2204.12983}{{\tt arXiv:2204.12983}}].

\bibitem{Chestnov:2022xsy}
V.~Chestnov, H.~Frellesvig, F.~Gasparotto, M.~K. Mandal, and P.~Mastrolia, {\it {Intersection Numbers from Higher-order Partial Differential Equations}},  \href{http://arxiv.org/abs/2209.01997}{{\tt arXiv:2209.01997}}.

\bibitem{Caron-Huot:2021xqj}
S.~Caron-Huot and A.~Pokraka, {\it {Duals of Feynman integrals. Part I. Differential equations}},  {\em JHEP} {\bf 12} (2021) 045, [\href{http://arxiv.org/abs/2104.06898}{{\tt arXiv:2104.06898}}].

\bibitem{Caron-Huot:2021iev}
S.~Caron-Huot and A.~Pokraka, {\it {Duals of Feynman Integrals. Part II. Generalized unitarity}},  {\em JHEP} {\bf 04} (2022) 078, [\href{http://arxiv.org/abs/2112.00055}{{\tt arXiv:2112.00055}}].

\bibitem{Weinzierl:2022eaz}
S.~Weinzierl, {\em {Feynman Integrals}}.
\newblock 1, 2022.

\bibitem{Cacciatori:2021nli}
S.~L. Cacciatori, M.~Conti, and S.~Trevisan, {\it {Co-Homology of Differential Forms and Feynman Diagrams}},  {\em Universe} {\bf 7} (2021), no.~9 328, [\href{http://arxiv.org/abs/2107.14721}{{\tt arXiv:2107.14721}}].

\bibitem{Cacciatori:2022mbi}
S.~L. Cacciatori and P.~Mastrolia, {\it {Intersection Numbers in Quantum Mechanics and Field Theory}},  \href{http://arxiv.org/abs/2211.03729}{{\tt arXiv:2211.03729}}.

\bibitem{Chen:2020uyk}
J.~Chen, X.~Jiang, X.~Xu, and L.~L. Yang, {\it {Constructing canonical Feynman integrals with intersection theory}},  {\em Phys. Lett. B} {\bf 814} (2021) 136085, [\href{http://arxiv.org/abs/2008.03045}{{\tt arXiv:2008.03045}}].

\bibitem{Chen:2022lzr}
J.~Chen, X.~Jiang, C.~Ma, X.~Xu, and L.~L. Yang, {\it {Baikov representations, intersection theory, and canonical Feynman integrals}},  {\em JHEP} {\bf 07} (2022) 066, [\href{http://arxiv.org/abs/2202.08127}{{\tt arXiv:2202.08127}}].

\bibitem{amslaurea27132}
G.~Fontana, {\it Rational algorithms for the decomposition of feynman integrals via intersection theory},  Master's thesis, 2022.

\bibitem{Baikov:1996iu}
P.~A. Baikov, {\it {Explicit solutions of the multiloop integral recurrence relations and its application}},  {\em Nucl. Instrum. Meth. A} {\bf 389} (1997) 347--349, [\href{http://arxiv.org/abs/hep-ph/9611449}{{\tt hep-ph/9611449}}].

\bibitem{Lee:2013hzt}
R.~N. Lee and A.~A. Pomeransky, {\it {Critical points and number of master integrals}},  {\em JHEP} {\bf 11} (2013) 165, [\href{http://arxiv.org/abs/1308.6676}{{\tt arXiv:1308.6676}}].

\bibitem{DeLaurentis:2022otd}
G.~De~Laurentis and B.~Page, {\it {Ans\"atze for scattering amplitudes from p-adic numbers and algebraic geometry}},  {\em JHEP} {\bf 12} (2022) 140, [\href{http://arxiv.org/abs/2203.04269}{{\tt arXiv:2203.04269}}].

\end{thebibliography}\endgroup
\end{document}